\documentclass[aps,prl,twocolumn,amsmath,amssymb,footinbib,showpacs,longbibliography,superscriptaddress]{revtex4-2}

\newcommand{\Jnature}{Nature (London)}
\newcommand{\Jnatphys}{Nat. Phys.}

\newcommand{\Jnatcomm}{Nat. Comm.}

\newcommand{\Jscience}{Science}

\newcommand{\Jpnas}{Proc. Nat. Acad.  Sci.}

\newcommand{\Jprl}{Phys. Rev. Lett.}

\newcommand{\Jprb}{Phys. Rev. B}

\newcommand{\Jepl}{Europhys. Lett.}

\newcommand{\crasphy}{C. R. Phys.}

\usepackage{amsmath}
\usepackage{amssymb}
\usepackage{graphicx}
\usepackage{float}
\usepackage[english]{babel} 

\usepackage[usenames,dvipsnames]{xcolor}
\usepackage[colorlinks=true,citecolor=Cerulean,linkcolor=RubineRed,urlcolor=Cerulean]{hyperref}

\newcommand{\ket}[1]{|{#1}\rangle}

\newcommand{\deanComment}[1]{\textcolor{blue}{#1}}

\newcommand{\kB}{k_{\textrm{\tiny B}}}
\newcommand{\Er}{E_{\textrm{r}}}
\newcommand{\fs}{f_{\textrm{s}}}

\begin{document}

\title{Weak Superfluidity in Twisted Optical Potentials}

\author{Dean Johnstone}
\affiliation{CPHT, CNRS, Ecole Polytechnique, IP Paris, F-91128 Palaiseau, France}

\author{Shanya Mishra}
\affiliation{CPHT, CNRS, Ecole Polytechnique, IP Paris, F-91128 Palaiseau, France}

\author{Zhaoxuan Zhu}
\affiliation{CPHT, CNRS, Ecole Polytechnique, IP Paris, F-91128 Palaiseau, France}

\author{Hepeng Yao}
\affiliation{DQMP, University of Geneva, 24 Quai Ernest-Ansermet, CH-1211 Geneva, Switzerland}

\author{Laurent Sanchez-Palencia}
\affiliation{CPHT, CNRS, Ecole Polytechnique, IP Paris, F-91128 Palaiseau, France}

\date{\today}

\begin{abstract}
A controlled twist between different underlying lattices allows one to interpolate, under a unified framework, across ordered and (quasi-)disordered matter while drastically changing quantum transport properties. Here, we use quantum Monte Carlo simulations to determine the unique phase diagrams of strongly-correlated ultracold bosons in twisted optical potentials. 
We show that at commensurate twisting angles, spectral gaps govern the formation of insulating patterns, separated by thin superfluid domains. The latter form weak superfluids, which are very sensitive to thermal fluctuations, but can be stabilized under appropriate parameter control. In contrast, slightly changing the twisting angle to a incommensurate value destroys most spectral gaps, leaving behind a prominent Bose glass phase. Our results are directly applicable to current generation experiments that quantum simulate moir\'e physics.
\end{abstract}

\maketitle

In the field of condensed matter, strongly correlated materials exhibit exotic phase transitions and entanglement properties. This behaviour may be enhanced in systems that have flat energy bands, such as stacked 2D materials, for they exaggerate the role of interactions over kinetic energy. To that end, much attention has been devoted to the novel superconducting and insulating phases of twisted bilayer graphene~\cite{cao2018unconventional,cao2018correlated,Yankowitz2019}, which forms flat bands at a set of magic twisting angles~\cite{Morell2010,Bistritzer2011,Tarnopolsky2019,Po2018,Gonzalez2019,Xie2020,choi2019electronic}. The resulting lattice has an enlarged unit cell, forming a so-called moir\'e, superlattice pattern~\cite{Santos2007,Santos2012,Zou2018}.
Away from magic angles, twisted models exhibit long-range quasiperiodic order, hence connecting to the physics quasicrystals~\cite{shechtman1984,steuer2004,steurer2018,kamiya2018,ahn2018,yao2018b}.
This case is, however, unlike in solid state bilayers, where commensurability via interactions is strongly favoured during fabrication~\cite{Santos2007,Hass2008,Kim2013}.
Emulating twisted models is also possible within the realm of quantum simulation, in particular with ultracold atoms, which offers high control over interactions and the underlying geometry of optical potentials~\cite{Lewenstein2007,Bloch2008,gross2017quantum,Dutta_2015,esslinger2010fermi,tarruell2018}.
Ultracold atoms in simple periodic potentials have been extensively studied and used to demonstrate Mott-insulator (MI) to superfluid (SF) phase transitions~\cite{Jaksch1998,greiner2002,jordens2008,schneider2008}.
If disorder is present, the Bose glass (BG), a special kind of compressible insulator, can also appear~\cite{giamarchi1987,giamarchi1988,Fisher1989,Kisker1997}, and has been observed in both disordered~\cite{Damski2003,lugan2007a,lugan2007b,Gurarie2009,Soyler2011,Carleo2013,lsp2010} and quasiperiodic models~\cite{Damski2003,Fallani2007,viebahn2019,sbroscia2020,Yao2020,Gautier2021,Johnstone_2022,Johnstone_2021,Zhu2023,JrChiunYu2023}. In recent years, these procedures have also been extended to twisted optical lattices \cite{cirac2019,Salamon2020,Luo2021,meng2023atomic,Junhyun2022}.
Here, the twist angle can be freely tuned between commensurate and incommensurate values.
So far, the role of incommensuration has been discussed in connection with ergodicity breaking in single-particle dynamics~\cite{paul2023particle} and SF-MI transitions in one-dimensional interacting Bose gases~\cite{yao2024}.

For this Letter, we study the exotic phase diagrams of strongly-correlated ultracold bosons within a twisted optical lattice, for both commensurate and incommensurate angles.
Using a combination of quantum Monte-Carlo (QMC) and exact diagonalisation, we show that in spite of the potentials having a very similar character, both cases behave completely differently. For incommensurate angles, our results confirm the presence of MI, SF, and BG phases, which we find to be closely linked to the localisation properties of single-particle states. For commensurate angles, the BG is replaced by a series of density-wave (DW) domains separated by weak SF regions, i.e.~a SF very unstable against thermal fluctuations. Such domains leave behind a normal fluid (NF)-like phase for even rather small temperatures.
We, however, show that control over the parameters in ultracold atom systems allows one to stabilize weak SF phases, which can be observed in current generation experiments. 
  
\textit{Model.--} We consider an interacting, 2D gas of ultracold bosons in an $L \times L$ box, with periodic boundary conditions. The Hamiltonian is
\begin{equation}\label{eq_mbh}
\hat{H} = \int d\mathbf{r} \,\, \hat{\Psi}^\dagger \left[ \hbar^2/2M \left(-\nabla^2 + \tilde{g}_0\hat{\Psi}^\dagger \hat{\Psi}\right) + V(\mathbf{r}) \right] \hat{\Psi},
\end{equation}
where $\hat{\Psi}$ is the bosonic field operator at point $\mathbf{r}=(x,y)$, $M$ is the atomic mass, $\tilde{g}_0$ is the dimensionless 2D contact interaction strength~\cite{Petrov2000,Petrov2001,Pricoupenko_2007,Ha2013}, and $V(\mathbf{r})$ is the optical potential.
The latter is constructed via two rotated square lattices, with period $a$,
\begin{equation} \label{eq_moireL}
V(\mathbf{r}) = V_1 v(R^{-} \mathbf{r}) + V_2 v(R^{+} \mathbf{r}),
\end{equation}
where
$v(\mathbf{r}) = \cos^2 (2\pi x/a) + \cos^2 (2\pi y/a)$,
$V_{1,2}$ are the potential depths,
and $R^{\pm}$ the rotation with angle $\pm \theta/2$.
Without loss of generality, we consider cases where $V_1=V_2=V$~\cite{note:V1V2}.

\begin{figure}[t!]
	\centering
	\makebox[0pt]{\includegraphics[width=0.99\linewidth]{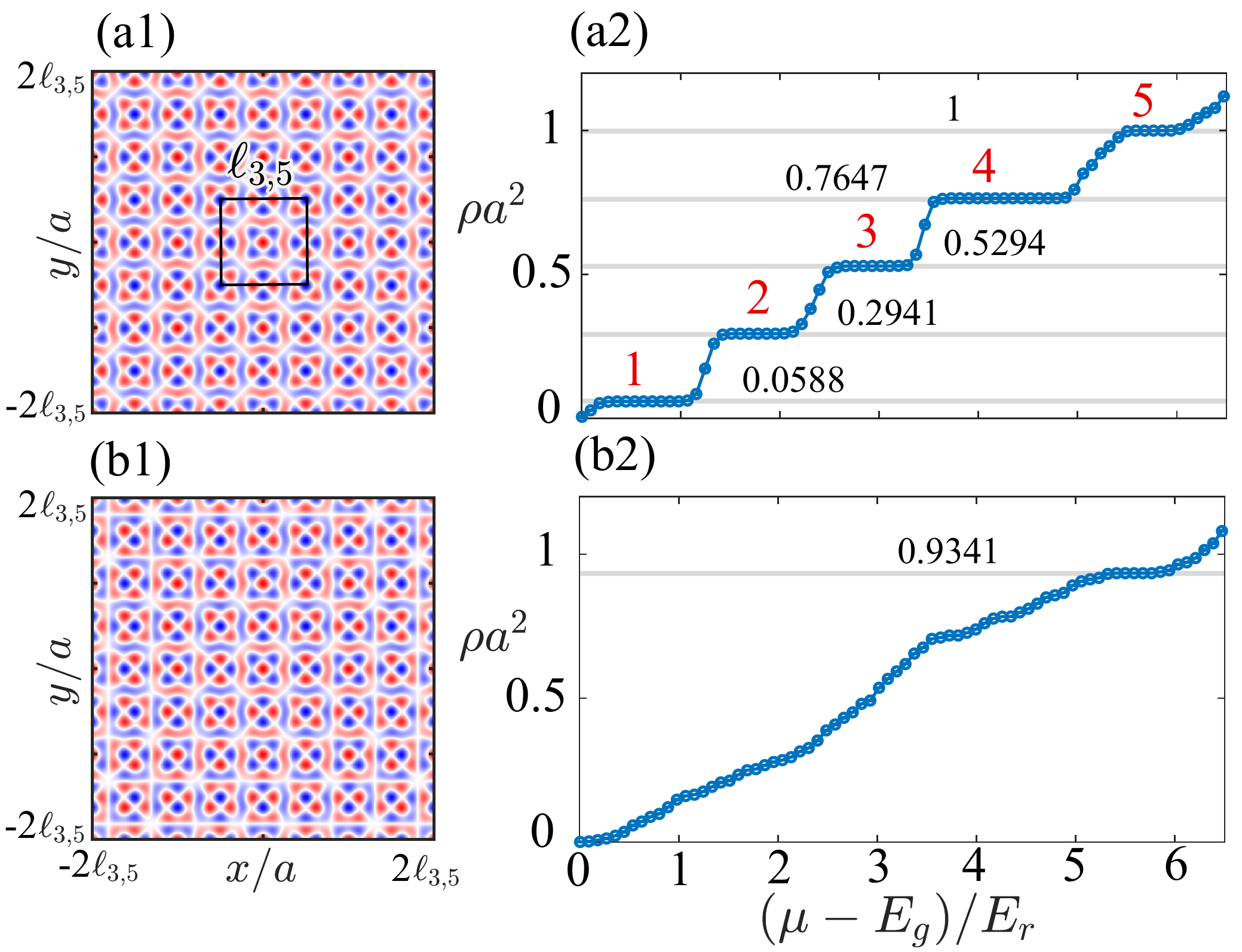}}
	\caption{(a1)~Commensurate potential with $\theta_{3,5} \approx 28.07^{\circ}$ (black square shows the unit cell) and (b1)~incommensurate potential with $\theta \approx 27.13^{\circ}$. (a2)-(b2)~QMC results for density versus shifted chemical potential at each respective angle for $V=6\Er$, $\tilde{g}_0=4$, $\kB T = 0.04\Er$ and $L=5\ell_{3,5}\approx20.62 a$. Plateaus in density are marked by grey lines, with $\rho a^2 < 1$~(DW) and $\rho a^2 = 1$~(MI). Red numbers in~(a2) denote the index of the plateau $p$, which has $1 + 4(p-1)$ atoms per unit cell.}
	\label{fig_density}
\end{figure}

All energies will be expressed in terms of the recoil energy $\Er=\pi^2 \hbar^2/2Ma^2$.
For almost any angle $\theta$, $V(\mathbf{r})$ forms an incommensurate (quasiperiodic) potential.
However, for a countable set of commensurate angles $\theta_{m,n}$, it takes the form of a commensurate (periodic) moir\'e potential.
These angles are defined by two coprime integers $m$ and $n$ \cite{cirac2019}, with
$\theta_{m,n} = \cos^{-1}\left(\frac{2mn}{m^2+n^2}\right)$,
where the length $\ell_{m,n}$
of the square unit cell is given by
\begin{equation}	\label{eq_c_length}
\ell_{m,n} = \begin{cases}
    a \sqrt{(m^2 + n^2)/2},& \textrm{if } m+n\, \textrm{even},\\
	a \sqrt{m^2 + n^2},& \textrm{if } m+n \, \textrm{odd}.
\end{cases}
\end{equation}
In Fig.~\ref{fig_density}, we compare
(a1)~a commensurate potential for the angle $\theta_{3,5}\approx 28.07^\circ$
with
(a2)~an incommensurate potential for a small shift in angle to $\theta = 27.13^{\circ}$.
The former is strictly periodic, with period $\ell_{3,5} \simeq 4.12 a$, and exhibits an exotic, decorated unit cell.
In contrast, the latter is not periodic, but shows a very similar potential profile, with hardly visible distortions.
These small differences, however, result in markedly distinct quantum phase diagrams.

\textit{Many-Body Phases.--}
To calculate the exact phases, we sample the many-body Hilbert space of Eq.~\eqref{eq_mbh} using worm QMC simulations in continuous space within the grand canonical ensemble~\cite{Boninsegni2006,Boninsegni2006b}, at finite temperature $T$ and chemical potential $\mu$. This gives us access to several order parameters
such as the average density $\rho=N/L^2$,
where $N$ is the total number of particles, the compressibility
$\kappa = {\partial \rho}/{\partial \mu}$,
and the SF fraction $\fs = {\Upsilon}/{\rho}$, where $\Upsilon$ is the superfluid stiffness~\cite{Ceperley1995}.
In Fig.~\ref{fig_density}, we plot the density $\rho$
versus the shifted chemical potential $\mu-E_g$ ($E_g$ is the single-particle ground state energy) for (a2)~commensurate and (b2)~incommensurate potentials. We fix $V=6\Er$, and consider a strongly interacting gas
with $\tilde{g}_0=4$.
Roughly speaking, ultracold atom experiments can reach temperatures of about $10\,$nK \deanComment{\cite{Li2020,Fabbri2011,Fabbri2015}}. For $^{87}$Rb atoms and a typical $a=350\,$nm, it gives $\kB T\approx0.04 \Er$, which we use in the QMC calculations for Fig.~\ref{fig_density}.

\begin{figure*}[t!]
	\centering
	\makebox[0pt]{\includegraphics[width=0.99\linewidth]{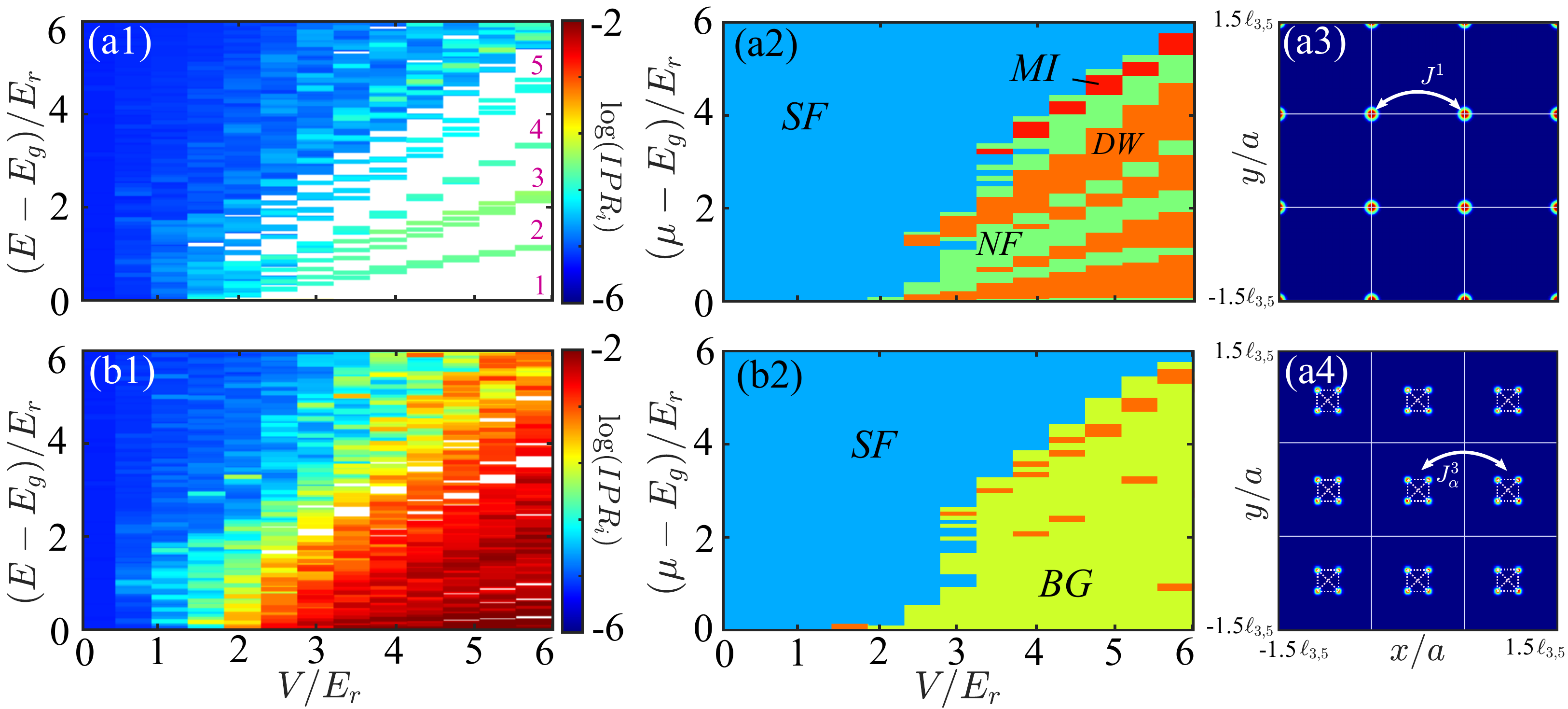}}
	\caption{(a1)-(b1)~$\textrm{IPR}_i$ of single-particle states for $L=15\ell_{3,5}\approx 61.85a$, at (a1)~commensurate angle with $\theta_{3,5} \approx 28.07^{\circ}$ and (b1)~incommensurate angle with $\theta \approx 27.13^{\circ}$.
(a2)-(b2)~QMC phase diagrams for the same potentials when $\tilde{g}_0=4$, $\kB T = 0.01\Er$, and $L=5\ell_{3,5}\approx20.62a$.
(a3)-(a4)~Examples of single-particle states below gap (a3)~1 and (a4)~3 are shown for the commensurate potential, which can be mapped to tight-binding models.
$J^{1}$ and $J^{\beta}_\alpha$ are couplings between states in nearest unit cells (white squares) and different bands and subbands (see text).
}
	\label{fig_sp}
\end{figure*}

First, for the commensurate case with $V=6\Er$ in Fig.~\ref{fig_density}(a2), $5$ plateaus in density (red numbers) are observed.
These correspond to incompressible, insulating phases with $\kappa=0$, in which we consistently find $\fs=0$ (not shown).
If the dimensionless density $\rho a^2$ is an integer, we have the standard MI phase. Here, we also find non-integer $\rho a^2$ plateaus, indicating a DW phase. The latter is similar in character to those identified in both superlattice~\cite{PhysRevB.73.174516,PhysRevA.81.053608} and quasiperiodic~\cite{Johnstone_2022,PhysRevA.78.023628} tight-binding models. Furthermore, a single DW-like phase has also been observed in experiments and mean-field calculations at $\theta_{1,22}\approx 5.21^{\circ}$ in Ref.~\cite{meng2023atomic}.
The increments in density are directly related to the geometry of the moir\'e potential.
First, the lowest DW plateau ($p=1$) fills the lowest potential minima [blue points at the $4$ corners of the unit cell in Fig.~\ref{fig_density}(a1)], giving $1$ atom per unit cell. The next DW plateau ($p=2$) fills the second $4$ lowest minima (located
near the midpoints of the unit cell edges), giving now $1+4=5$ atoms per unit cell.
This pattern continues for all insulating phases with $\rho a^2\le1$.
The density for the $p$-th plateau is then $\rho_p = [1+4(p-1)]/\ell_{m,n}^2$, corresponding to the black numbers in Figs~\ref{fig_density}(a2), (a3),
and the total number of plateaus is $\mathcal{M}_{m,n} = 1 + (\ell^2_{m,n}/a^2-1)/4$.
For the commensurate potential, we therefore have $\ell^2_{3,5}=17a^2$ and $\mathcal{M}_{3,5}=5$, consistently with the results of Fig.~\ref{fig_density}(a2).

Figure~\ref{fig_density}(b2) shows the counterpart of Fig.~\ref{fig_density}(a2) for the incommensurate potential.
In spite of the strong similarity between the potentials in Figs.~\ref{fig_density}(a1) and (b1), most plateaus are now absent.
To explain this, we note that any incommensurate potential can be approximated by a commensurate one with very large $m$ and $n$.
This implies that $\mathcal{M}_{m,n}$ diverges, so that each plateau becomes vanishingly narrow. The spectral gaps then vanish, and the compressibility is finite.
Moreover, the moir\'e period increases, which weakens phase coherence and induces localisation in the incommensurate limit.
This forms a BG, with $\kappa>0$ and $\fs=0$.
Note, that a small DW plateau with $\rho a^2\approx0.93$ survives, due to a small single-particle gap in the quasi-periodic potential.

\textit{Weak Superfluids.--} In our calculations, we find $\fs=0$ for both potentials, consistently with the onset of DW, MI, and BG phases. 
However, $\fs=0$ is also found for the commensurate potential in compressible domains, while a SF is to be expected due to the underlying periodicity of the system. The absence of finite $\fs$ may be attributed to the moir\'e-enhanced lattice period and the finite temperature. In other words, tunnelling is vanishingly small, and the system is thus unable to stabilise superfluidity, even against very weak thermal fluctuations. 

To check this, we compare QMC phase diagrams with single-particle calculations.
Figures.~\ref{fig_sp}(a1) and (b1) show the single-particle spectra versus energy and potential depth
for both commensurate and incommensurate potentials, with $L=15\ell_{3,5}\approx61.85a$.
The spectra are coloured according to the inverse participation ratio $\textrm{IPR}_i=\int d\mathbf{r} | \psi_i(\mathbf{r}) |^4$ of each state $i$, with blue (red) regions mapping to extended (localised and/or critical) states.
Starting with the incommensurate case in Fig.~\ref{fig_sp}(b1), the low energy states are primarily localised above a critical potential depth (here $V\approx2.5\Er$), as expected for a quasi-periodic system.
The onset of single-particle localisation is both qualitatively and quantitatively consistent with the BG phase for the corresponding many-body phase diagram in Fig.~\ref{fig_sp}(b2), while extended states map onto the SF phase.
Narrow DW regions, consistent with single-particle gaps, are also visible. 

Let us now turn to the commensurate case. The single- particle spectrum in Fig.~\ref{fig_sp}(a1) displays $5$ prominent gaps (labelled 1-5), which quantitatively overlap the $5$ insulating plateaus in the many-body phase diagram of Fig.~\ref{fig_sp}(a2) (1-4 are DWs, 5 is the MI). This quantitative mapping is to be expected, owing to 2D fermionization of strongly-repelling bosons. Here, the single-particle states in different bands occupy separated regions of the plane, so that energy is minimised by filling each with up to one boson, hence mimicking local Pauli exclusion~\cite{Zhu2023,Zhu2024}.
The spatial separation of single-particle states is illustrated in Figs.~\ref{fig_sp}(a3) and (a4), which show the density distribution of eigenstates in the first and third band, respectively.
In the first band, the density is concentrated at the nodes of the moir\'e Bravais lattice (white lines), while in the third band, it is concentrated around four points near the centre of each unit cell.
Similar separation occurs with the other bands.

Due to periodicity and Bloch's theorem, all states are extended, with the underlying density profiles in Figs.~\ref{fig_sp}(a3) and (a4) implying that different bands with index $\beta$ correspond to different tight-binding geometries~\cite{note:SupplMat}.
For band~$\beta=1$, the picture is straightforward since there is $1$ site per unit cell, see Fig.~\ref{fig_sp}(a3). This yields a standard square lattice with period $\ell_{3,5}$, which has a  dispersion relation of
\begin{equation} \label{eq_k_sq}
\varepsilon(\mathbf{k})= \epsilon^1 -2\big[J^1_x \cos(k_x \ell_{3,5}) + J^1_y \cos(k_y \ell_{3,5})\big],
\end{equation}
where $J^1_x=J^1_y=J^1$ owing to 4-fold rotation symmetry.
Here $J^1$ is the tunnelling rate, $\epsilon^1$ is the onsite energy, and $\mathbf{k}=(k_x,\, k_y)$ spans the $1$st Brillouin zone $[-\pi/\ell_{3,5} \dots +\pi/\ell_{3,5}]$.
For $V \gtrsim 3\Er$, we find that the tight-binding prediction of Eq.~\eqref{eq_k_sq} is in excellent agreement with the continuous space bandstructure.

For the other bands, we have a different scenario in which each moir\'e cell contains 4 sites, see for instance Fig.~\ref{fig_sp}(a4) corresponding to band~3.
Intra-cell couplings dominate over inter-cell couplings, so that each band splits into 4 subbands $\alpha$ (=$a$, $b$, $c$, and $d$) with width much smaller than the total band width.
The subbands $a$ and $d$ are well separated in energy, and can thus described by a standard tight-binding model, yielding dispersion relations similar to Eq.~(\ref{eq_k_sq}) with $J^\beta_{a,d; x}=J^\beta_{a,d; y}=J^\beta_{a,d}$.
In contrast, the subbands $b$ and $c$ are quasi-degenerate and the 4-fold rotation symmetry is broken due to inter-cell coupling, i.e.~$J^\beta_{b,c;x} \neq J^\beta_{b,c;y}$~\cite{note:rt}.

By fitting these dispersion relations to the continuous space bandstructure, we can then extract all relevant tunnelling energies.
For $V=6\Er$, we find that the largest tunnelling in each band, $J^\beta_{\textrm{\tiny max}}=\max_\alpha(\vert J^\beta_\alpha\vert)$ ranges from $J^1_{\textrm{\tiny max}} \sim 10^{-11}\Er$ to $J^5_{\textrm{\tiny max}} \sim 10^{-3} \Er$~\cite{note:SupplMat}.

To observe a distinct SF phase within band $\beta$, we typically require that $\kB T \lesssim J^\beta_{\textrm{\tiny max}}$~\cite{Gerbier2007}.
For $V=6\Er$, it means $\kB  T \lesssim 0.001\Er$, i.e.~for $^{87}$Rb atoms $T \lesssim 0.25\,$nK, quite unrealistic for present-day experiments.
For QMC simulations and typical experiments with $\kB  T \approx 0.04 \Er$, thermal broadening
strongly depletes the SF domains, leaving behind a thermal NF phase, which explains the absence of observable superfluidity in the calculations of Fig.~\ref{fig_density}.
This effect can, however, be somewhat mitigated by considering smaller potential depths, in which the effective couplings $J^{\beta}_\alpha$ then become larger.
For instance, when $V=3\Er$, the largest tunnellings culminate to $J^{2}_{\textrm{\tiny max}} \approx 0.0134\Er$, within band $2$.
QMC calculations for varying temperatures confirm that a sizeable SF fraction appears roughly for $\kB T \sim J^{2}_{\textrm{\tiny max}}$~\cite{note:SupplMat}.
The phase diagrams in Fig.~\ref{fig_sp} are plotted for $\kB T=0.01\Er$, slightly below $J^{2}_{\textrm{\tiny max}}$ and, consistently, a small SF domain appears for $V=3\Er$ and $\mu-E_g \sim 1.5\Er$ in Fig.~\ref{fig_sp}(a2). Such temperatures may be reached using different atomic species, e.g.~$^{40}$K and $^{7}$Li, for which $\kB T \approx 0.02 \Er$ and $\kB T \approx 0.0036\Er$, respectively, when $T=10$\,nK.
Due to the small values of $J^\beta_\alpha$, however, SF domains between insulators are very narrow and only present around $V=3\Er$, with thermal fluctuations leading to the prevalence of the NF.

\begin{figure}[t!]
	\centering
	\makebox[0pt]{\includegraphics[width=0.85\linewidth]{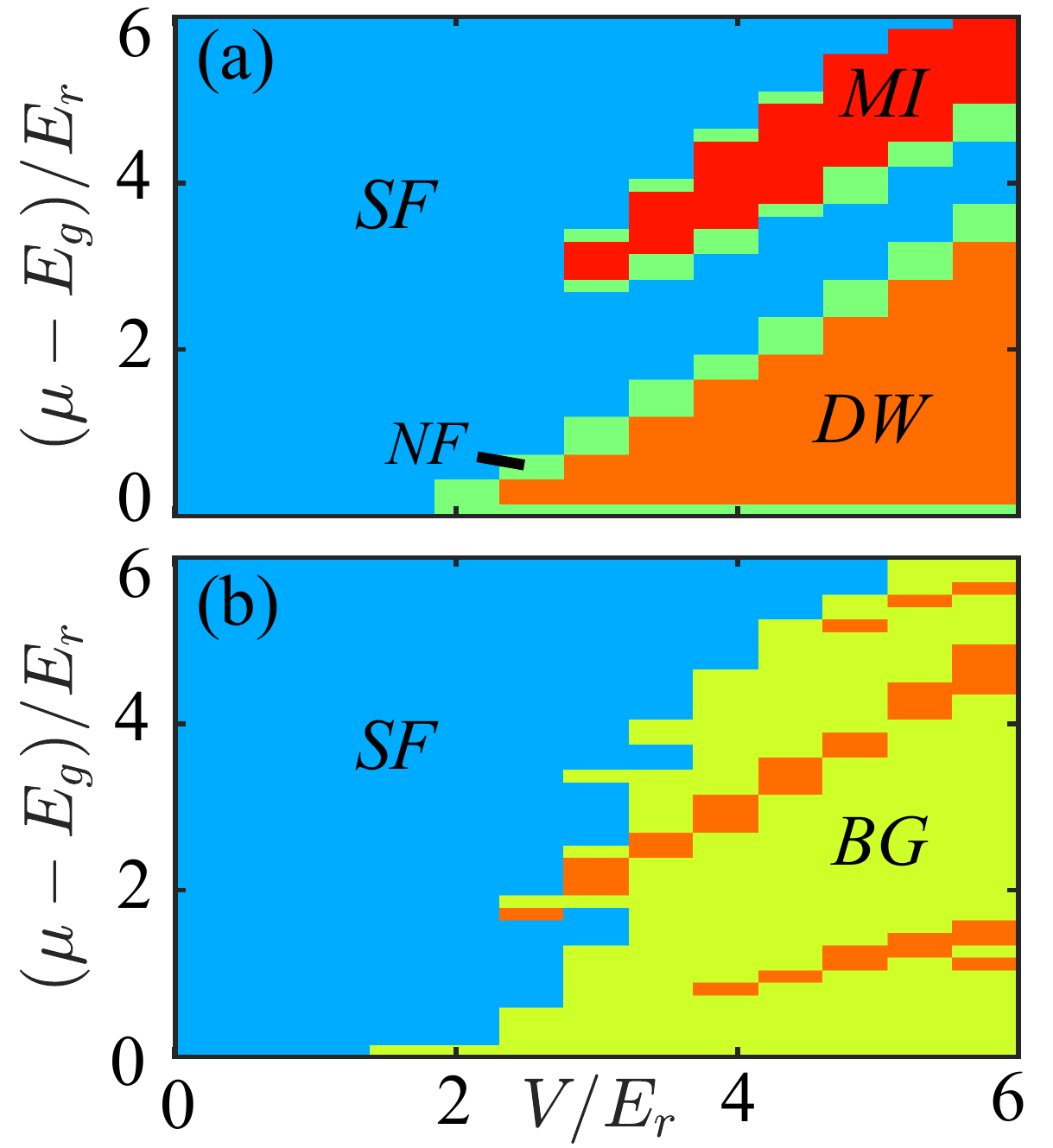}}
	\caption{QMC phase diagrams with $\tilde{g}_0=4$, $\kB T = 0.01\Er$ and $L=10\ell_{2,1}\approx22.36a$, for the (a)~commensurate angle $\theta_{2,1}\approx36.87^{\circ}$ and (b)~incommensurate angle $\theta\approx37.82^{\circ}$.
	}
	\label{fig_phases}
\end{figure}

To further enhance superfluidity between the insulating plateaus without further reducing $T$ or $V$, we may exploit another controllable parameter of ultracold atoms, namely the twist angle $\theta$.
In Fig.~\ref{fig_phases}(a), we plot another phase diagram for the same parameters as in Fig.~\ref{fig_sp}(a2), but with $\theta=\theta_{2,1}\approx36.87^{\circ}$.
At this commensurate angle, the size of the unit cell is now $\ell_{2,1}\approx2.24a$, giving $\mathcal{M}_{2,1}=2$.
We consistently find a single DW lobe (with $\rho a^2 = 0.2$) below the MI (with $\rho a^2 = 1$).
The smaller unit cell enhances tunnelling rates, which now culminate to $J^{2}_{\textrm{max}}=0.042\Er$ for $V=6\Er$.
Then, for a temperature $T =0.01\Er/\kB$ (sufficiently below $J^{2}_{\textrm{max}}$), we find a prominent SF phase, leaving behind narrow NF domains in the vicinity of the edges of the DW and MI lobes. 
In between the DW and MI lobes, we find a significant SF fraction, which varies from $\fs \approx 0.2$ at $V=3\Er$ to $\fs \approx0.1$ at $V=6\Er$.
Consistently with the results above, a drastic change in behaviour is observed for a small angular variation to the incommensurate $\theta\approx37.82^{\circ}$ in Fig.~\ref{fig_phases}(b). Large DW and MI lobes vanish, as well as the intermediate SF, leaving behind a BG that is qualitatively similar to that observed for $\theta \approx 27.13^\circ$ in Fig.~\ref{fig_sp}(b2).

\textit{Conclusions.--} We have shown that ultracold bosons in twisted optical potentials with controlled twist angle can undergo a variety of exotic phase transitions. Incommensurate angles induce quasiperiodic potentials, which support BG phases over a large ranges of potential depths and chemical potentials.
On the other hand, commensurate angles create moir\'e potentials, which can stabilise a family of DW lobes.
Such insulating domains map onto single-particle gaps, which separate bands and narrow subbands.
The latter support weak SF domains,
which are highly sensitive to finite temperature effects.
Sizeable SF domains can, however, be stabilized by controlling the potential depth and, more importantly, the twist angle.
The twisted potentials we considered are realisable with the current generation of ultracold atom experiments,
using a configurations similar to that used in Refs.~\cite{viebahn2019,sbroscia2020,JrChiunYu2023}, and the various phases can be identified using now standard matter-wave interferometry and transport measurements.
An interesting extension of our work would be to consider twisted bilayer lattices, where the two layers are associated with different internal atomic states~\cite{cirac2019}. Such systems allow for the simulation of twisted bilayer materials in condensed matter. In ultracold atom simulators, independent control of the inter-layer and intra-layer couplings, as well as the twist angle, offer a further means to stabilize new exotic phases~\cite{meng2023atomic}.

\begin{acknowledgments}
We thank Jing Zhang and Shengjie Yu for fruitful discussions.
We acknowledge the CPHT computer team for valuable support.
This research was supported by
the Agence Nationale de la Recherche (ANR, project ANR-CMAQ-002 France~2030),
the program `Investissements d'Avenir'', the LabEx PALM (project ANR-10-LABX-0039-PALM),
the IPParis Doctoral School, the Swiss National Science Foundation (Gant numbers 200020-188687 and 200020-219400),
and HPC/AI resources from GENCI-TGCC (Grant 2023-A0110510300) using the ALPS scheduler library and statistical analysis tools \cite{troyer1998,ALPS2011}.
\end{acknowledgments}

\newpage


\begin{thebibliography}{78}%
	\makeatletter
	\providecommand \@ifxundefined [1]{%
		\@ifx{#1\undefined}
	}%
	\providecommand \@ifnum [1]{%
		\ifnum #1\expandafter \@firstoftwo
		\else \expandafter \@secondoftwo
		\fi
	}%
	\providecommand \@ifx [1]{%
		\ifx #1\expandafter \@firstoftwo
		\else \expandafter \@secondoftwo
		\fi
	}%
	\providecommand \natexlab [1]{#1}%
	\providecommand \enquote  [1]{``#1''}%
	\providecommand \bibnamefont  [1]{#1}%
	\providecommand \bibfnamefont [1]{#1}%
	\providecommand \citenamefont [1]{#1}%
	\providecommand \href@noop [0]{\@secondoftwo}%
	\providecommand \href [0]{\begingroup \@sanitize@url \@href}%
	\providecommand \@href[1]{\@@startlink{#1}\@@href}%
	\providecommand \@@href[1]{\endgroup#1\@@endlink}%
	\providecommand \@sanitize@url [0]{\catcode `\\12\catcode `\$12\catcode
		`\&12\catcode `\#12\catcode `\^12\catcode `\_12\catcode `\%12\relax}%
	\providecommand \@@startlink[1]{}%
	\providecommand \@@endlink[0]{}%
	\providecommand \url  [0]{\begingroup\@sanitize@url \@url }%
	\providecommand \@url [1]{\endgroup\@href {#1}{\urlprefix }}%
	\providecommand \urlprefix  [0]{URL }%
	\providecommand \Eprint [0]{\href }%
	\providecommand \doibase [0]{https://doi.org/}%
	\providecommand \selectlanguage [0]{\@gobble}%
	\providecommand \bibinfo  [0]{\@secondoftwo}%
	\providecommand \bibfield  [0]{\@secondoftwo}%
	\providecommand \translation [1]{[#1]}%
	\providecommand \BibitemOpen [0]{}%
	\providecommand \bibitemStop [0]{}%
	\providecommand \bibitemNoStop [0]{.\EOS\space}%
	\providecommand \EOS [0]{\spacefactor3000\relax}%
	\providecommand \BibitemShut  [1]{\csname bibitem#1\endcsname}%
	\let\auto@bib@innerbib\@empty
	\bibitem [{\citenamefont {Cao}\ \emph {et~al.}(2018{\natexlab{a}})\citenamefont
		{Cao}, \citenamefont {Fatemi}, \citenamefont {Fang}, \citenamefont
		{Watanabe}, \citenamefont {Taniguchi}, \citenamefont {Kaxiras},\ and\
		\citenamefont {Jarillo-Herrero}}]{cao2018unconventional}%
	\BibitemOpen
	\bibfield  {author} {\bibinfo {author} {\bibfnamefont {Y.}~\bibnamefont
			{Cao}}, \bibinfo {author} {\bibfnamefont {V.}~\bibnamefont {Fatemi}},
		\bibinfo {author} {\bibfnamefont {S.}~\bibnamefont {Fang}}, \bibinfo {author}
		{\bibfnamefont {K.}~\bibnamefont {Watanabe}}, \bibinfo {author}
		{\bibfnamefont {T.}~\bibnamefont {Taniguchi}}, \bibinfo {author}
		{\bibfnamefont {E.}~\bibnamefont {Kaxiras}},\ and\ \bibinfo {author}
		{\bibfnamefont {P.}~\bibnamefont {Jarillo-Herrero}},\ }\bibfield  {title}
	{\bibinfo {title} {Unconventional superconductivity in magic-angle graphene
			superlattices},\ }\href {https://doi.org/10.1038/nature26160} {\bibfield
		{journal} {\bibinfo  {journal} {Nature}\ }\textbf {\bibinfo {volume} {556}},\
		\bibinfo {pages} {43} (\bibinfo {year} {2018}{\natexlab{a}})}\BibitemShut
	{NoStop}%
	\bibitem [{\citenamefont {Cao}\ \emph {et~al.}(2018{\natexlab{b}})\citenamefont
		{Cao}, \citenamefont {Fatemi}, \citenamefont {Demir}, \citenamefont {Fang},
		\citenamefont {Tomarken}, \citenamefont {Luo}, \citenamefont
		{Sanchez-Yamagishi}, \citenamefont {Watanabe}, \citenamefont {Taniguchi},
		\citenamefont {Kaxiras} \emph {et~al.}}]{cao2018correlated}%
	\BibitemOpen
	\bibfield  {author} {\bibinfo {author} {\bibfnamefont {Y.}~\bibnamefont
			{Cao}}, \bibinfo {author} {\bibfnamefont {V.}~\bibnamefont {Fatemi}},
		\bibinfo {author} {\bibfnamefont {A.}~\bibnamefont {Demir}}, \bibinfo
		{author} {\bibfnamefont {S.}~\bibnamefont {Fang}}, \bibinfo {author}
		{\bibfnamefont {S.~L.}\ \bibnamefont {Tomarken}}, \bibinfo {author}
		{\bibfnamefont {J.~Y.}\ \bibnamefont {Luo}}, \bibinfo {author} {\bibfnamefont
			{J.~D.}\ \bibnamefont {Sanchez-Yamagishi}}, \bibinfo {author} {\bibfnamefont
			{K.}~\bibnamefont {Watanabe}}, \bibinfo {author} {\bibfnamefont
			{T.}~\bibnamefont {Taniguchi}}, \bibinfo {author} {\bibfnamefont
			{E.}~\bibnamefont {Kaxiras}}, \emph {et~al.},\ }\bibfield  {title} {\bibinfo
		{title} {Correlated insulator behaviour at half-filling in magic-angle
			graphene superlattices},\ }\href {https://doi.org/10.1038/nature26154}
	{\bibfield  {journal} {\bibinfo  {journal} {Nature}\ }\textbf {\bibinfo
			{volume} {556}},\ \bibinfo {pages} {80} (\bibinfo {year}
		{2018}{\natexlab{b}})}\BibitemShut {NoStop}%
	\bibitem [{\citenamefont {Yankowitz}\ \emph {et~al.}(2019)\citenamefont
		{Yankowitz}, \citenamefont {Chen}, \citenamefont {Polshyn}, \citenamefont
		{Zhang}, \citenamefont {Watanabe}, \citenamefont {Taniguchi}, \citenamefont
		{Graf}, \citenamefont {Young},\ and\ \citenamefont {Dean}}]{Yankowitz2019}%
	\BibitemOpen
	\bibfield  {author} {\bibinfo {author} {\bibfnamefont {M.}~\bibnamefont
			{Yankowitz}}, \bibinfo {author} {\bibfnamefont {S.}~\bibnamefont {Chen}},
		\bibinfo {author} {\bibfnamefont {H.}~\bibnamefont {Polshyn}}, \bibinfo
		{author} {\bibfnamefont {Y.}~\bibnamefont {Zhang}}, \bibinfo {author}
		{\bibfnamefont {K.}~\bibnamefont {Watanabe}}, \bibinfo {author}
		{\bibfnamefont {T.}~\bibnamefont {Taniguchi}}, \bibinfo {author}
		{\bibfnamefont {D.}~\bibnamefont {Graf}}, \bibinfo {author} {\bibfnamefont
			{A.~F.}\ \bibnamefont {Young}},\ and\ \bibinfo {author} {\bibfnamefont
			{C.~R.}\ \bibnamefont {Dean}},\ }\bibfield  {title} {\bibinfo {title} {Tuning
			superconductivity in twisted bilayer graphene},\ }\href
	{https://doi.org/10.1126/science.aav1910} {\bibfield  {journal} {\bibinfo
			{journal} {Science}\ }\textbf {\bibinfo {volume} {363}},\ \bibinfo {pages}
		{1059} (\bibinfo {year} {2019})}\BibitemShut {NoStop}%
	\bibitem [{\citenamefont {Su\'arez~Morell}\ \emph {et~al.}(2010)\citenamefont
		{Su\'arez~Morell}, \citenamefont {Correa}, \citenamefont {Vargas},
		\citenamefont {Pacheco},\ and\ \citenamefont {Barticevic}}]{Morell2010}%
	\BibitemOpen
	\bibfield  {author} {\bibinfo {author} {\bibfnamefont {E.}~\bibnamefont
			{Su\'arez~Morell}}, \bibinfo {author} {\bibfnamefont {J.~D.}\ \bibnamefont
			{Correa}}, \bibinfo {author} {\bibfnamefont {P.}~\bibnamefont {Vargas}},
		\bibinfo {author} {\bibfnamefont {M.}~\bibnamefont {Pacheco}},\ and\ \bibinfo
		{author} {\bibfnamefont {Z.}~\bibnamefont {Barticevic}},\ }\bibfield  {title}
	{\bibinfo {title} {Flat bands in slightly twisted bilayer graphene:
			Tight-binding calculations},\ }\href
	{https://doi.org/10.1103/PhysRevB.82.121407} {\bibfield  {journal} {\bibinfo
			{journal} {Phys. Rev. B}\ }\textbf {\bibinfo {volume} {82}},\ \bibinfo
		{pages} {121407} (\bibinfo {year} {2010})}\BibitemShut {NoStop}%
	\bibitem [{\citenamefont {Bistritzer}\ and\ \citenamefont
		{MacDonald}(2011)}]{Bistritzer2011}%
	\BibitemOpen
	\bibfield  {author} {\bibinfo {author} {\bibfnamefont {R.}~\bibnamefont
			{Bistritzer}}\ and\ \bibinfo {author} {\bibfnamefont {A.~H.}\ \bibnamefont
			{MacDonald}},\ }\bibfield  {title} {\bibinfo {title} {Moiré bands in twisted
			double-layer graphene},\ }\href {https://doi.org/10.1073/pnas.1108174108}
	{\bibfield  {journal} {\bibinfo  {journal} {Proc. Natl. Acad. Sci.}\ }\textbf
		{\bibinfo {volume} {108}},\ \bibinfo {pages} {12233} (\bibinfo {year}
		{2011})}\BibitemShut {NoStop}%
	\bibitem [{\citenamefont {Tarnopolsky}\ \emph {et~al.}(2019)\citenamefont
		{Tarnopolsky}, \citenamefont {Kruchkov},\ and\ \citenamefont
		{Vishwanath}}]{Tarnopolsky2019}%
	\BibitemOpen
	\bibfield  {author} {\bibinfo {author} {\bibfnamefont {G.}~\bibnamefont
			{Tarnopolsky}}, \bibinfo {author} {\bibfnamefont {A.~J.}\ \bibnamefont
			{Kruchkov}},\ and\ \bibinfo {author} {\bibfnamefont {A.}~\bibnamefont
			{Vishwanath}},\ }\bibfield  {title} {\bibinfo {title} {Origin of magic angles
			in twisted bilayer graphene},\ }\href
	{https://doi.org/10.1103/PhysRevLett.122.106405} {\bibfield  {journal}
		{\bibinfo  {journal} {Phys. Rev. Lett.}\ }\textbf {\bibinfo {volume} {122}},\
		\bibinfo {pages} {106405} (\bibinfo {year} {2019})}\BibitemShut {NoStop}%
	\bibitem [{\citenamefont {Po}\ \emph {et~al.}(2018)\citenamefont {Po},
		\citenamefont {Zou}, \citenamefont {Vishwanath},\ and\ \citenamefont
		{Senthil}}]{Po2018}%
	\BibitemOpen
	\bibfield  {author} {\bibinfo {author} {\bibfnamefont {H.~C.}\ \bibnamefont
			{Po}}, \bibinfo {author} {\bibfnamefont {L.}~\bibnamefont {Zou}}, \bibinfo
		{author} {\bibfnamefont {A.}~\bibnamefont {Vishwanath}},\ and\ \bibinfo
		{author} {\bibfnamefont {T.}~\bibnamefont {Senthil}},\ }\bibfield  {title}
	{\bibinfo {title} {Origin of mott insulating behavior and superconductivity
			in twisted bilayer graphene},\ }\href
	{https://doi.org/10.1103/PhysRevX.8.031089} {\bibfield  {journal} {\bibinfo
			{journal} {Phys. Rev. X}\ }\textbf {\bibinfo {volume} {8}},\ \bibinfo {pages}
		{031089} (\bibinfo {year} {2018})}\BibitemShut {NoStop}%
	\bibitem [{\citenamefont {Gonz\'alez}\ and\ \citenamefont
		{Stauber}(2019)}]{Gonzalez2019}%
	\BibitemOpen
	\bibfield  {author} {\bibinfo {author} {\bibfnamefont {J.}~\bibnamefont
			{Gonz\'alez}}\ and\ \bibinfo {author} {\bibfnamefont {T.}~\bibnamefont
			{Stauber}},\ }\bibfield  {title} {\bibinfo {title} {Kohn-luttinger
			superconductivity in twisted bilayer graphene},\ }\href
	{https://doi.org/10.1103/PhysRevLett.122.026801} {\bibfield  {journal}
		{\bibinfo  {journal} {Phys. Rev. Lett.}\ }\textbf {\bibinfo {volume} {122}},\
		\bibinfo {pages} {026801} (\bibinfo {year} {2019})}\BibitemShut {NoStop}%
	\bibitem [{\citenamefont {Xie}\ and\ \citenamefont
		{MacDonald}(2020)}]{Xie2020}%
	\BibitemOpen
	\bibfield  {author} {\bibinfo {author} {\bibfnamefont {M.}~\bibnamefont
			{Xie}}\ and\ \bibinfo {author} {\bibfnamefont {A.~H.}\ \bibnamefont
			{MacDonald}},\ }\bibfield  {title} {\bibinfo {title} {Nature of the
			correlated insulator states in twisted bilayer graphene},\ }\href
	{https://doi.org/10.1103/PhysRevLett.124.097601} {\bibfield  {journal}
		{\bibinfo  {journal} {Phys. Rev. Lett.}\ }\textbf {\bibinfo {volume} {124}},\
		\bibinfo {pages} {097601} (\bibinfo {year} {2020})}\BibitemShut {NoStop}%
	\bibitem [{\citenamefont {Choi}\ \emph {et~al.}(2019)\citenamefont {Choi},
		\citenamefont {Kemmer}, \citenamefont {Peng}, \citenamefont {Thomson},
		\citenamefont {Arora}, \citenamefont {Polski}, \citenamefont {Zhang},
		\citenamefont {Ren}, \citenamefont {Alicea}, \citenamefont {Refael} \emph
		{et~al.}}]{choi2019electronic}%
	\BibitemOpen
	\bibfield  {author} {\bibinfo {author} {\bibfnamefont {Y.}~\bibnamefont
			{Choi}}, \bibinfo {author} {\bibfnamefont {J.}~\bibnamefont {Kemmer}},
		\bibinfo {author} {\bibfnamefont {Y.}~\bibnamefont {Peng}}, \bibinfo {author}
		{\bibfnamefont {A.}~\bibnamefont {Thomson}}, \bibinfo {author} {\bibfnamefont
			{H.}~\bibnamefont {Arora}}, \bibinfo {author} {\bibfnamefont
			{R.}~\bibnamefont {Polski}}, \bibinfo {author} {\bibfnamefont
			{Y.}~\bibnamefont {Zhang}}, \bibinfo {author} {\bibfnamefont
			{H.}~\bibnamefont {Ren}}, \bibinfo {author} {\bibfnamefont {J.}~\bibnamefont
			{Alicea}}, \bibinfo {author} {\bibfnamefont {G.}~\bibnamefont {Refael}},
		\emph {et~al.},\ }\bibfield  {title} {\bibinfo {title} {Electronic
			correlations in twisted bilayer graphene near the magic angle},\ }\href
	{https://doi.org/10.1038/s41567-019-0606-5} {\bibfield  {journal} {\bibinfo
			{journal} {Nat. Phys.}\ }\textbf {\bibinfo {volume} {15}},\ \bibinfo {pages}
		{1174} (\bibinfo {year} {2019})}\BibitemShut {NoStop}%
	\bibitem [{\citenamefont {Lopes~dos Santos}\ \emph {et~al.}(2007)\citenamefont
		{Lopes~dos Santos}, \citenamefont {Peres},\ and\ \citenamefont
		{Castro~Neto}}]{Santos2007}%
	\BibitemOpen
	\bibfield  {author} {\bibinfo {author} {\bibfnamefont {J.~M.~B.}\
			\bibnamefont {Lopes~dos Santos}}, \bibinfo {author} {\bibfnamefont
			{N.~M.~R.}\ \bibnamefont {Peres}},\ and\ \bibinfo {author} {\bibfnamefont
			{A.~H.}\ \bibnamefont {Castro~Neto}},\ }\bibfield  {title} {\bibinfo {title}
		{Graphene bilayer with a twist: Electronic structure},\ }\href
	{https://doi.org/10.1103/PhysRevLett.99.256802} {\bibfield  {journal}
		{\bibinfo  {journal} {Phys. Rev. Lett.}\ }\textbf {\bibinfo {volume} {99}},\
		\bibinfo {pages} {256802} (\bibinfo {year} {2007})}\BibitemShut {NoStop}%
	\bibitem [{\citenamefont {Lopes~dos Santos}\ \emph {et~al.}(2012)\citenamefont
		{Lopes~dos Santos}, \citenamefont {Peres},\ and\ \citenamefont
		{Castro~Neto}}]{Santos2012}%
	\BibitemOpen
	\bibfield  {author} {\bibinfo {author} {\bibfnamefont {J.~M.~B.}\
			\bibnamefont {Lopes~dos Santos}}, \bibinfo {author} {\bibfnamefont
			{N.~M.~R.}\ \bibnamefont {Peres}},\ and\ \bibinfo {author} {\bibfnamefont
			{A.~H.}\ \bibnamefont {Castro~Neto}},\ }\bibfield  {title} {\bibinfo {title}
		{Continuum model of the twisted graphene bilayer},\ }\href
	{https://doi.org/10.1103/PhysRevB.86.155449} {\bibfield  {journal} {\bibinfo
			{journal} {Phys. Rev. B}\ }\textbf {\bibinfo {volume} {86}},\ \bibinfo
		{pages} {155449} (\bibinfo {year} {2012})}\BibitemShut {NoStop}%
	\bibitem [{\citenamefont {Zou}\ \emph {et~al.}(2018)\citenamefont {Zou},
		\citenamefont {Po}, \citenamefont {Vishwanath},\ and\ \citenamefont
		{Senthil}}]{Zou2018}%
	\BibitemOpen
	\bibfield  {author} {\bibinfo {author} {\bibfnamefont {L.}~\bibnamefont
			{Zou}}, \bibinfo {author} {\bibfnamefont {H.~C.}\ \bibnamefont {Po}},
		\bibinfo {author} {\bibfnamefont {A.}~\bibnamefont {Vishwanath}},\ and\
		\bibinfo {author} {\bibfnamefont {T.}~\bibnamefont {Senthil}},\ }\bibfield
	{title} {\bibinfo {title} {Band structure of twisted bilayer graphene:
			Emergent symmetries, commensurate approximants, and wannier obstructions},\
	}\href {https://doi.org/10.1103/PhysRevB.98.085435} {\bibfield  {journal}
		{\bibinfo  {journal} {Phys. Rev. B}\ }\textbf {\bibinfo {volume} {98}},\
		\bibinfo {pages} {085435} (\bibinfo {year} {2018})}\BibitemShut {NoStop}%
	\bibitem [{\citenamefont {Shechtman}\ \emph {et~al.}(1984)\citenamefont
		{Shechtman}, \citenamefont {Blech}, \citenamefont {Gratias},\ and\
		\citenamefont {Cahn}}]{shechtman1984}%
	\BibitemOpen
	\bibfield  {author} {\bibinfo {author} {\bibfnamefont {D.}~\bibnamefont
			{Shechtman}}, \bibinfo {author} {\bibfnamefont {I.}~\bibnamefont {Blech}},
		\bibinfo {author} {\bibfnamefont {D.}~\bibnamefont {Gratias}},\ and\ \bibinfo
		{author} {\bibfnamefont {J.~W.}\ \bibnamefont {Cahn}},\ }\bibfield  {title}
	{\bibinfo {title} {Metallic phase with long-range orientational order and no
			translational symmetry},\ }\href@noop {} {\bibfield  {journal} {\bibinfo
			{journal} {\Jprl}\ }\textbf {\bibinfo {volume} {53}},\ \bibinfo {pages}
		{1951} (\bibinfo {year} {1984})}\BibitemShut {NoStop}%
	\bibitem [{\citenamefont {Steurer}(2004)}]{steuer2004}%
	\BibitemOpen
	\bibfield  {author} {\bibinfo {author} {\bibfnamefont {W.}~\bibnamefont
			{Steurer}},\ }\bibfield  {title} {\bibinfo {title} {Twenty years of structure
			research on quasicrystals. {P}art~{I}.~{P}entagonal, octagonal, decagonal and
			dodecagonal quasicrystals},\ }\href@noop {} {\bibfield  {journal} {\bibinfo
			{journal} {Z. Kristallogr. Cryst. Mater.}\ }\textbf {\bibinfo {volume}
			{219}},\ \bibinfo {pages} {391} (\bibinfo {year} {2004})}\BibitemShut
	{NoStop}%
	\bibitem [{\citenamefont {Steurer}(2018)}]{steurer2018}%
	\BibitemOpen
	\bibfield  {author} {\bibinfo {author} {\bibfnamefont {W.}~\bibnamefont
			{Steurer}},\ }\bibfield  {title} {\bibinfo {title} {{Quasicrystals: {W}hat do
				we know? {W}hat do we want to know? {W}hat can we know?}},\ }\href@noop {}
	{\bibfield  {journal} {\bibinfo  {journal} {Acta Cryst. A}\ }\textbf
		{\bibinfo {volume} {74}},\ \bibinfo {pages} {1} (\bibinfo {year}
		{2018})}\BibitemShut {NoStop}%
	\bibitem [{\citenamefont {Kamiya}\ \emph {et~al.}(2018)\citenamefont {Kamiya},
		\citenamefont {Takeuchi}, \citenamefont {Kabeya}, \citenamefont {Wada},
		\citenamefont {Ishimasa}, \citenamefont {Ochiai}, \citenamefont {Deguchi},
		\citenamefont {Imura},\ and\ \citenamefont {Sato}}]{kamiya2018}%
	\BibitemOpen
	\bibfield  {author} {\bibinfo {author} {\bibfnamefont {K.}~\bibnamefont
			{Kamiya}}, \bibinfo {author} {\bibfnamefont {T.}~\bibnamefont {Takeuchi}},
		\bibinfo {author} {\bibfnamefont {N.}~\bibnamefont {Kabeya}}, \bibinfo
		{author} {\bibfnamefont {N.}~\bibnamefont {Wada}}, \bibinfo {author}
		{\bibfnamefont {T.}~\bibnamefont {Ishimasa}}, \bibinfo {author}
		{\bibfnamefont {A.}~\bibnamefont {Ochiai}}, \bibinfo {author} {\bibfnamefont
			{K.}~\bibnamefont {Deguchi}}, \bibinfo {author} {\bibfnamefont
			{K.}~\bibnamefont {Imura}},\ and\ \bibinfo {author} {\bibfnamefont
			{N.}~\bibnamefont {Sato}},\ }\bibfield  {title} {\bibinfo {title} {Discovery
			of superconductivity in quasicrystal},\ }\href@noop {} {\bibfield  {journal}
		{\bibinfo  {journal} {\Jnatcomm}\ }\textbf {\bibinfo {volume} {9}},\ \bibinfo
		{pages} {1} (\bibinfo {year} {2018})}\BibitemShut {NoStop}%
	\bibitem [{\citenamefont {Ahn}\ \emph {et~al.}(2018)\citenamefont {Ahn},
		\citenamefont {Moon}, \citenamefont {Kim}, \citenamefont {Kim}, \citenamefont
		{Shin}, \citenamefont {Kim}, \citenamefont {Cha}, \citenamefont {Kahng},
		\citenamefont {Kim}, \citenamefont {Koshino}, \citenamefont {Son},
		\citenamefont {Yang},\ and\ \citenamefont {Ahn}}]{ahn2018}%
	\BibitemOpen
	\bibfield  {author} {\bibinfo {author} {\bibfnamefont {S.~J.}\ \bibnamefont
			{Ahn}}, \bibinfo {author} {\bibfnamefont {P.}~\bibnamefont {Moon}}, \bibinfo
		{author} {\bibfnamefont {T.-H.}\ \bibnamefont {Kim}}, \bibinfo {author}
		{\bibfnamefont {H.-W.}\ \bibnamefont {Kim}}, \bibinfo {author} {\bibfnamefont
			{H.-C.}\ \bibnamefont {Shin}}, \bibinfo {author} {\bibfnamefont {E.~H.}\
			\bibnamefont {Kim}}, \bibinfo {author} {\bibfnamefont {H.~W.}\ \bibnamefont
			{Cha}}, \bibinfo {author} {\bibfnamefont {S.-J.}\ \bibnamefont {Kahng}},
		\bibinfo {author} {\bibfnamefont {P.}~\bibnamefont {Kim}}, \bibinfo {author}
		{\bibfnamefont {M.}~\bibnamefont {Koshino}}, \bibinfo {author} {\bibfnamefont
			{Y.-W.}\ \bibnamefont {Son}}, \bibinfo {author} {\bibfnamefont {C.-W.}\
			\bibnamefont {Yang}},\ and\ \bibinfo {author} {\bibfnamefont {J.~R.}\
			\bibnamefont {Ahn}},\ }\bibfield  {title} {\bibinfo {title} {Dirac electrons
			in a dodecagonal graphene quasicrystal},\ }\href@noop {} {\bibfield
		{journal} {\bibinfo  {journal} {\Jscience}\ }\textbf {\bibinfo {volume}
			{361}},\ \bibinfo {pages} {782} (\bibinfo {year} {2018})}\BibitemShut
	{NoStop}%
	\bibitem [{\citenamefont {Yao}\ \emph {et~al.}(2018)\citenamefont {Yao},
		\citenamefont {Wang}, \citenamefont {Bao}, \citenamefont {Zhang},
		\citenamefont {Zhang}, \citenamefont {Bao}, \citenamefont {Chan},
		\citenamefont {Chen}, \citenamefont {Avila}, \citenamefont {Asensio},
		\citenamefont {Zhu},\ and\ \citenamefont {Zhou}}]{yao2018b}%
	\BibitemOpen
	\bibfield  {author} {\bibinfo {author} {\bibfnamefont {W.}~\bibnamefont
			{Yao}}, \bibinfo {author} {\bibfnamefont {E.}~\bibnamefont {Wang}}, \bibinfo
		{author} {\bibfnamefont {C.}~\bibnamefont {Bao}}, \bibinfo {author}
		{\bibfnamefont {Y.}~\bibnamefont {Zhang}}, \bibinfo {author} {\bibfnamefont
			{K.}~\bibnamefont {Zhang}}, \bibinfo {author} {\bibfnamefont
			{K.}~\bibnamefont {Bao}}, \bibinfo {author} {\bibfnamefont {C.~K.}\
			\bibnamefont {Chan}}, \bibinfo {author} {\bibfnamefont {C.}~\bibnamefont
			{Chen}}, \bibinfo {author} {\bibfnamefont {J.}~\bibnamefont {Avila}},
		\bibinfo {author} {\bibfnamefont {M.~C.}\ \bibnamefont {Asensio}}, \bibinfo
		{author} {\bibfnamefont {J.}~\bibnamefont {Zhu}},\ and\ \bibinfo {author}
		{\bibfnamefont {S.}~\bibnamefont {Zhou}},\ }\bibfield  {title} {\bibinfo
		{title} {Quasicrystalline 30{\textdegree} twisted bilayer graphene as an
			incommensurate superlattice with strong interlayer coupling},\ }\href@noop {}
	{\bibfield  {journal} {\bibinfo  {journal} {\Jpnas}\ }\textbf {\bibinfo
			{volume} {115}},\ \bibinfo {pages} {6928} (\bibinfo {year}
		{2018})}\BibitemShut {NoStop}%
	\bibitem [{\citenamefont {Hass}\ \emph {et~al.}(2008)\citenamefont {Hass},
		\citenamefont {Varchon}, \citenamefont {Mill\'an-Otoya}, \citenamefont
		{Sprinkle}, \citenamefont {Sharma}, \citenamefont {de~Heer}, \citenamefont
		{Berger}, \citenamefont {First}, \citenamefont {Magaud},\ and\ \citenamefont
		{Conrad}}]{Hass2008}%
	\BibitemOpen
	\bibfield  {author} {\bibinfo {author} {\bibfnamefont {J.}~\bibnamefont
			{Hass}}, \bibinfo {author} {\bibfnamefont {F.}~\bibnamefont {Varchon}},
		\bibinfo {author} {\bibfnamefont {J.~E.}\ \bibnamefont {Mill\'an-Otoya}},
		\bibinfo {author} {\bibfnamefont {M.}~\bibnamefont {Sprinkle}}, \bibinfo
		{author} {\bibfnamefont {N.}~\bibnamefont {Sharma}}, \bibinfo {author}
		{\bibfnamefont {W.~A.}\ \bibnamefont {de~Heer}}, \bibinfo {author}
		{\bibfnamefont {C.}~\bibnamefont {Berger}}, \bibinfo {author} {\bibfnamefont
			{P.~N.}\ \bibnamefont {First}}, \bibinfo {author} {\bibfnamefont
			{L.}~\bibnamefont {Magaud}},\ and\ \bibinfo {author} {\bibfnamefont {E.~H.}\
			\bibnamefont {Conrad}},\ }\bibfield  {title} {\bibinfo {title} {Why
			multilayer graphene on
			$4h\mathrm{\text{\ensuremath{-}}}\mathrm{SiC}(000\overline{1})$ behaves like
			a single sheet of graphene},\ }\href
	{https://doi.org/10.1103/PhysRevLett.100.125504} {\bibfield  {journal}
		{\bibinfo  {journal} {Phys. Rev. Lett.}\ }\textbf {\bibinfo {volume} {100}},\
		\bibinfo {pages} {125504} (\bibinfo {year} {2008})}\BibitemShut {NoStop}%
	\bibitem [{\citenamefont {Kim}\ \emph {et~al.}(2013)\citenamefont {Kim},
		\citenamefont {Yun}, \citenamefont {Nam}, \citenamefont {Son}, \citenamefont
		{Lee}, \citenamefont {Kim}, \citenamefont {Seo}, \citenamefont {Choi},
		\citenamefont {Lee}, \citenamefont {Lee},\ and\ \citenamefont
		{Kim}}]{Kim2013}%
	\BibitemOpen
	\bibfield  {author} {\bibinfo {author} {\bibfnamefont {Y.}~\bibnamefont
			{Kim}}, \bibinfo {author} {\bibfnamefont {H.}~\bibnamefont {Yun}}, \bibinfo
		{author} {\bibfnamefont {S.-G.}\ \bibnamefont {Nam}}, \bibinfo {author}
		{\bibfnamefont {M.}~\bibnamefont {Son}}, \bibinfo {author} {\bibfnamefont
			{D.~S.}\ \bibnamefont {Lee}}, \bibinfo {author} {\bibfnamefont {D.~C.}\
			\bibnamefont {Kim}}, \bibinfo {author} {\bibfnamefont {S.}~\bibnamefont
			{Seo}}, \bibinfo {author} {\bibfnamefont {H.~C.}\ \bibnamefont {Choi}},
		\bibinfo {author} {\bibfnamefont {H.-J.}\ \bibnamefont {Lee}}, \bibinfo
		{author} {\bibfnamefont {S.~W.}\ \bibnamefont {Lee}},\ and\ \bibinfo {author}
		{\bibfnamefont {J.~S.}\ \bibnamefont {Kim}},\ }\bibfield  {title} {\bibinfo
		{title} {Breakdown of the interlayer coherence in twisted bilayer graphene},\
	}\href {https://doi.org/10.1103/PhysRevLett.110.096602} {\bibfield  {journal}
		{\bibinfo  {journal} {Phys. Rev. Lett.}\ }\textbf {\bibinfo {volume} {110}},\
		\bibinfo {pages} {096602} (\bibinfo {year} {2013})}\BibitemShut {NoStop}%
	\bibitem [{\citenamefont {Lewenstein}\ \emph {et~al.}(2007)\citenamefont
		{Lewenstein}, \citenamefont {Sanpera}, \citenamefont {Ahufinger},
		\citenamefont {Damski}, \citenamefont {Sen(De)},\ and\ \citenamefont
		{Sen}}]{Lewenstein2007}%
	\BibitemOpen
	\bibfield  {author} {\bibinfo {author} {\bibfnamefont {M.}~\bibnamefont
			{Lewenstein}}, \bibinfo {author} {\bibfnamefont {A.}~\bibnamefont {Sanpera}},
		\bibinfo {author} {\bibfnamefont {V.}~\bibnamefont {Ahufinger}}, \bibinfo
		{author} {\bibfnamefont {B.}~\bibnamefont {Damski}}, \bibinfo {author}
		{\bibfnamefont {A.}~\bibnamefont {Sen(De)}},\ and\ \bibinfo {author}
		{\bibfnamefont {U.}~\bibnamefont {Sen}},\ }\bibfield  {title} {\bibinfo
		{title} {Ultracold atomic gases in optical lattices: mimicking condensed
			matter physics and beyond},\ }\href
	{https://doi.org/10.1080/00018730701223200} {\bibfield  {journal} {\bibinfo
			{journal} {Adv. Phys.}\ }\textbf {\bibinfo {volume} {56}},\ \bibinfo {pages}
		{243} (\bibinfo {year} {2007})}\BibitemShut {NoStop}%
	\bibitem [{\citenamefont {Bloch}\ \emph {et~al.}(2008)\citenamefont {Bloch},
		\citenamefont {Dalibard},\ and\ \citenamefont {Zwerger}}]{Bloch2008}%
	\BibitemOpen
	\bibfield  {author} {\bibinfo {author} {\bibfnamefont {I.}~\bibnamefont
			{Bloch}}, \bibinfo {author} {\bibfnamefont {J.}~\bibnamefont {Dalibard}},\
		and\ \bibinfo {author} {\bibfnamefont {W.}~\bibnamefont {Zwerger}},\
	}\bibfield  {title} {\bibinfo {title} {Many-body physics with ultracold
			gases},\ }\href {https://doi.org/10.1103/RevModPhys.80.885} {\bibfield
		{journal} {\bibinfo  {journal} {Rev. Mod. Phys.}\ }\textbf {\bibinfo {volume}
			{80}},\ \bibinfo {pages} {885} (\bibinfo {year} {2008})}\BibitemShut
	{NoStop}%
	\bibitem [{\citenamefont {Gross}\ and\ \citenamefont
		{Bloch}(2017)}]{gross2017quantum}%
	\BibitemOpen
	\bibfield  {author} {\bibinfo {author} {\bibfnamefont {C.}~\bibnamefont
			{Gross}}\ and\ \bibinfo {author} {\bibfnamefont {I.}~\bibnamefont {Bloch}},\
	}\bibfield  {title} {\bibinfo {title} {Quantum simulations with ultracold
			atoms in optical lattices},\ }\href {https://doi.org/10.1126/science.aal3837}
	{\bibfield  {journal} {\bibinfo  {journal} {Science}\ }\textbf {\bibinfo
			{volume} {357}},\ \bibinfo {pages} {995} (\bibinfo {year}
		{2017})}\BibitemShut {NoStop}%
	\bibitem [{\citenamefont {Dutta}\ \emph {et~al.}(2015)\citenamefont {Dutta},
		\citenamefont {Gajda}, \citenamefont {Hauke}, \citenamefont {Lewenstein},
		\citenamefont {Lühmann}, \citenamefont {Malomed}, \citenamefont
		{Sowiński},\ and\ \citenamefont {Zakrzewski}}]{Dutta_2015}%
	\BibitemOpen
	\bibfield  {author} {\bibinfo {author} {\bibfnamefont {O.}~\bibnamefont
			{Dutta}}, \bibinfo {author} {\bibfnamefont {M.}~\bibnamefont {Gajda}},
		\bibinfo {author} {\bibfnamefont {P.}~\bibnamefont {Hauke}}, \bibinfo
		{author} {\bibfnamefont {M.}~\bibnamefont {Lewenstein}}, \bibinfo {author}
		{\bibfnamefont {D.-S.}\ \bibnamefont {Lühmann}}, \bibinfo {author}
		{\bibfnamefont {B.~A.}\ \bibnamefont {Malomed}}, \bibinfo {author}
		{\bibfnamefont {T.}~\bibnamefont {Sowiński}},\ and\ \bibinfo {author}
		{\bibfnamefont {J.}~\bibnamefont {Zakrzewski}},\ }\bibfield  {title}
	{\bibinfo {title} {Non-standard hubbard models in optical lattices: a
			review},\ }\href {https://doi.org/10.1088/0034-4885/78/6/066001} {\bibfield
		{journal} {\bibinfo  {journal} {Rep. Prog. Phys.}\ }\textbf {\bibinfo
			{volume} {78}},\ \bibinfo {pages} {066001} (\bibinfo {year}
		{2015})}\BibitemShut {NoStop}%
	\bibitem [{\citenamefont {Esslinger}(2010)}]{esslinger2010fermi}%
	\BibitemOpen
	\bibfield  {author} {\bibinfo {author} {\bibfnamefont {T.}~\bibnamefont
			{Esslinger}},\ }\bibfield  {title} {\bibinfo {title} {Fermi-hubbard physics
			with atoms in an optical lattice},\ }\href
	{https://doi.org/10.1146/annurev-conmatphys-070909-104059} {\bibfield
		{journal} {\bibinfo  {journal} {Annu. Rev. Condens. Matter Phys.}\ }\textbf
		{\bibinfo {volume} {1}},\ \bibinfo {pages} {129} (\bibinfo {year}
		{2010})}\BibitemShut {NoStop}%
	\bibitem [{\citenamefont {Tarruell}\ and\ \citenamefont
		{Sanchez-Palencia}(2018)}]{tarruell2018}%
	\BibitemOpen
	\bibfield  {author} {\bibinfo {author} {\bibfnamefont {L.}~\bibnamefont
			{Tarruell}}\ and\ \bibinfo {author} {\bibfnamefont {L.}~\bibnamefont
			{Sanchez-Palencia}},\ }\bibfield  {title} {\bibinfo {title} {Quantum
			simulation of the {H}ubbard model with ultracold fermions in optical
			lattices},\ }\href@noop {} {\bibfield  {journal} {\bibinfo  {journal}
			{\crasphy}\ }\textbf {\bibinfo {volume} {19}},\ \bibinfo {pages} {365}
		(\bibinfo {year} {2018})}\BibitemShut {NoStop}%
	\bibitem [{\citenamefont {Jaksch}\ \emph {et~al.}(1998)\citenamefont {Jaksch},
		\citenamefont {Bruder}, \citenamefont {Cirac}, \citenamefont {Gardiner},\
		and\ \citenamefont {Zoller}}]{Jaksch1998}%
	\BibitemOpen
	\bibfield  {author} {\bibinfo {author} {\bibfnamefont {D.}~\bibnamefont
			{Jaksch}}, \bibinfo {author} {\bibfnamefont {C.}~\bibnamefont {Bruder}},
		\bibinfo {author} {\bibfnamefont {J.~I.}\ \bibnamefont {Cirac}}, \bibinfo
		{author} {\bibfnamefont {C.~W.}\ \bibnamefont {Gardiner}},\ and\ \bibinfo
		{author} {\bibfnamefont {P.}~\bibnamefont {Zoller}},\ }\bibfield  {title}
	{\bibinfo {title} {Cold bosonic atoms in optical lattices},\ }\href
	{https://doi.org/10.1103/PhysRevLett.81.3108} {\bibfield  {journal} {\bibinfo
			{journal} {Phys. Rev. Lett.}\ }\textbf {\bibinfo {volume} {81}},\ \bibinfo
		{pages} {3108} (\bibinfo {year} {1998})}\BibitemShut {NoStop}%
	\bibitem [{\citenamefont {Greiner}\ \emph {et~al.}(2002)\citenamefont
		{Greiner}, \citenamefont {Mandel}, \citenamefont {Esslinger}, \citenamefont
		{H\"ansch},\ and\ \citenamefont {Bloch}}]{greiner2002}%
	\BibitemOpen
	\bibfield  {author} {\bibinfo {author} {\bibfnamefont {M.}~\bibnamefont
			{Greiner}}, \bibinfo {author} {\bibfnamefont {O.}~\bibnamefont {Mandel}},
		\bibinfo {author} {\bibfnamefont {T.}~\bibnamefont {Esslinger}}, \bibinfo
		{author} {\bibfnamefont {T.~W.}\ \bibnamefont {H\"ansch}},\ and\ \bibinfo
		{author} {\bibfnamefont {I.}~\bibnamefont {Bloch}},\ }\bibfield  {title}
	{\bibinfo {title} {Quantum phase transition from a superfluid to a {M}ott
			insulator in a gas of ultracold atoms},\ }\href@noop {} {\bibfield  {journal}
		{\bibinfo  {journal} {\Jnature}\ }\textbf {\bibinfo {volume} {415}},\
		\bibinfo {pages} {39} (\bibinfo {year} {2002})}\BibitemShut {NoStop}%
	\bibitem [{\citenamefont {J\"ordens}\ \emph {et~al.}(2008)\citenamefont
		{J\"ordens}, \citenamefont {Strohmaier}, \citenamefont {G\"unter},
		\citenamefont {Moritz},\ and\ \citenamefont {Esslinger}}]{jordens2008}%
	\BibitemOpen
	\bibfield  {author} {\bibinfo {author} {\bibfnamefont {R.}~\bibnamefont
			{J\"ordens}}, \bibinfo {author} {\bibfnamefont {N.}~\bibnamefont
			{Strohmaier}}, \bibinfo {author} {\bibfnamefont {K.}~\bibnamefont
			{G\"unter}}, \bibinfo {author} {\bibfnamefont {H.}~\bibnamefont {Moritz}},\
		and\ \bibinfo {author} {\bibfnamefont {T.}~\bibnamefont {Esslinger}},\
	}\bibfield  {title} {\bibinfo {title} {A {M}ott insulator of fermionic atoms
			in an optical lattice},\ }\href@noop {} {\bibfield  {journal} {\bibinfo
			{journal} {\Jnature}\ }\textbf {\bibinfo {volume} {455}},\ \bibinfo {pages}
		{204} (\bibinfo {year} {2008})}\BibitemShut {NoStop}%
	\bibitem [{\citenamefont {Schneider}\ \emph {et~al.}(2008)\citenamefont
		{Schneider}, \citenamefont {Hackerm\"uller}, \citenamefont {Will},
		\citenamefont {Best}, \citenamefont {Bloch}, \citenamefont {Costi},
		\citenamefont {Helmes}, \citenamefont {Rasch},\ and\ \citenamefont
		{Rosch}}]{schneider2008}%
	\BibitemOpen
	\bibfield  {author} {\bibinfo {author} {\bibfnamefont {U.}~\bibnamefont
			{Schneider}}, \bibinfo {author} {\bibfnamefont {L.}~\bibnamefont
			{Hackerm\"uller}}, \bibinfo {author} {\bibfnamefont {S.}~\bibnamefont
			{Will}}, \bibinfo {author} {\bibfnamefont {T.}~\bibnamefont {Best}}, \bibinfo
		{author} {\bibfnamefont {I.}~\bibnamefont {Bloch}}, \bibinfo {author}
		{\bibfnamefont {T.~A.}\ \bibnamefont {Costi}}, \bibinfo {author}
		{\bibfnamefont {R.~W.}\ \bibnamefont {Helmes}}, \bibinfo {author}
		{\bibfnamefont {D.}~\bibnamefont {Rasch}},\ and\ \bibinfo {author}
		{\bibfnamefont {A.}~\bibnamefont {Rosch}},\ }\bibfield  {title} {\bibinfo
		{title} {Metallic and insulating phases of repulsively interacting fermions
			in a 3{D} optical lattice},\ }\href@noop {} {\bibfield  {journal} {\bibinfo
			{journal} {\Jscience}\ }\textbf {\bibinfo {volume} {322}},\ \bibinfo {pages}
		{1520} (\bibinfo {year} {2008})}\BibitemShut {NoStop}%
	\bibitem [{\citenamefont {Giamarchi}\ and\ \citenamefont
		{Schulz}(1987)}]{giamarchi1987}%
	\BibitemOpen
	\bibfield  {author} {\bibinfo {author} {\bibfnamefont {T.}~\bibnamefont
			{Giamarchi}}\ and\ \bibinfo {author} {\bibfnamefont {H.~J.}\ \bibnamefont
			{Schulz}},\ }\bibfield  {title} {\bibinfo {title} {Localization and
			interactions in one-dimensional quantum fluids},\ }\href@noop {} {\bibfield
		{journal} {\bibinfo  {journal} {\Jepl}\ }\textbf {\bibinfo {volume} {3}},\
		\bibinfo {pages} {1287} (\bibinfo {year} {1987})}\BibitemShut {NoStop}%
	\bibitem [{\citenamefont {Giamarchi}\ and\ \citenamefont
		{Schulz}(1988)}]{giamarchi1988}%
	\BibitemOpen
	\bibfield  {author} {\bibinfo {author} {\bibfnamefont {T.}~\bibnamefont
			{Giamarchi}}\ and\ \bibinfo {author} {\bibfnamefont {H.~J.}\ \bibnamefont
			{Schulz}},\ }\bibfield  {title} {\bibinfo {title} {Anderson localization and
			interactions in one-dimensional metals},\ }\href@noop {} {\bibfield
		{journal} {\bibinfo  {journal} {\Jprb}\ }\textbf {\bibinfo {volume} {37}},\
		\bibinfo {pages} {325} (\bibinfo {year} {1988})}\BibitemShut {NoStop}%
	\bibitem [{\citenamefont {Fisher}\ \emph {et~al.}(1989)\citenamefont {Fisher},
		\citenamefont {Weichman}, \citenamefont {Grinstein},\ and\ \citenamefont
		{Fisher}}]{Fisher1989}%
	\BibitemOpen
	\bibfield  {author} {\bibinfo {author} {\bibfnamefont {M.~P.~A.}\
			\bibnamefont {Fisher}}, \bibinfo {author} {\bibfnamefont {P.~B.}\
			\bibnamefont {Weichman}}, \bibinfo {author} {\bibfnamefont {G.}~\bibnamefont
			{Grinstein}},\ and\ \bibinfo {author} {\bibfnamefont {D.~S.}\ \bibnamefont
			{Fisher}},\ }\bibfield  {title} {\bibinfo {title} {Boson localization and the
			superfluid-insulator transition},\ }\href
	{https://doi.org/10.1103/PhysRevB.40.546} {\bibfield  {journal} {\bibinfo
			{journal} {Phys. Rev. B}\ }\textbf {\bibinfo {volume} {40}},\ \bibinfo
		{pages} {546} (\bibinfo {year} {1989})}\BibitemShut {NoStop}%
	\bibitem [{\citenamefont {Kisker}\ and\ \citenamefont
		{Rieger}(1997)}]{Kisker1997}%
	\BibitemOpen
	\bibfield  {author} {\bibinfo {author} {\bibfnamefont {J.}~\bibnamefont
			{Kisker}}\ and\ \bibinfo {author} {\bibfnamefont {H.}~\bibnamefont
			{Rieger}},\ }\bibfield  {title} {\bibinfo {title} {Bose-glass and
			mott-insulator phase in the disordered boson hubbard model},\ }\href
	{https://doi.org/10.1103/PhysRevB.55.R11981} {\bibfield  {journal} {\bibinfo
			{journal} {Phys. Rev. B}\ }\textbf {\bibinfo {volume} {55}},\ \bibinfo
		{pages} {R11981} (\bibinfo {year} {1997})}\BibitemShut {NoStop}%
	\bibitem [{\citenamefont {Damski}\ \emph {et~al.}(2003)\citenamefont {Damski},
		\citenamefont {Zakrzewski}, \citenamefont {Santos}, \citenamefont {Zoller},\
		and\ \citenamefont {Lewenstein}}]{Damski2003}%
	\BibitemOpen
	\bibfield  {author} {\bibinfo {author} {\bibfnamefont {B.}~\bibnamefont
			{Damski}}, \bibinfo {author} {\bibfnamefont {J.}~\bibnamefont {Zakrzewski}},
		\bibinfo {author} {\bibfnamefont {L.}~\bibnamefont {Santos}}, \bibinfo
		{author} {\bibfnamefont {P.}~\bibnamefont {Zoller}},\ and\ \bibinfo {author}
		{\bibfnamefont {M.}~\bibnamefont {Lewenstein}},\ }\bibfield  {title}
	{\bibinfo {title} {Atomic bose and anderson glasses in optical lattices},\
	}\href {https://doi.org/10.1103/PhysRevLett.91.080403} {\bibfield  {journal}
		{\bibinfo  {journal} {Phys. Rev. Lett.}\ }\textbf {\bibinfo {volume} {91}},\
		\bibinfo {pages} {080403} (\bibinfo {year} {2003})}\BibitemShut {NoStop}%
	\bibitem [{\citenamefont {Lugan}\ \emph
		{et~al.}(2007{\natexlab{a}})\citenamefont {Lugan}, \citenamefont {Cl\'ement},
		\citenamefont {Bouyer}, \citenamefont {Aspect}, \citenamefont {Lewenstein},\
		and\ \citenamefont {Sanchez-Palencia}}]{lugan2007a}%
	\BibitemOpen
	\bibfield  {author} {\bibinfo {author} {\bibfnamefont {P.}~\bibnamefont
			{Lugan}}, \bibinfo {author} {\bibfnamefont {D.}~\bibnamefont {Cl\'ement}},
		\bibinfo {author} {\bibfnamefont {P.}~\bibnamefont {Bouyer}}, \bibinfo
		{author} {\bibfnamefont {A.}~\bibnamefont {Aspect}}, \bibinfo {author}
		{\bibfnamefont {M.}~\bibnamefont {Lewenstein}},\ and\ \bibinfo {author}
		{\bibfnamefont {L.}~\bibnamefont {Sanchez-Palencia}},\ }\bibfield  {title}
	{\bibinfo {title} {Ultracold {B}ose gases in 1{D} disorder: From {L}ifshits
			glass to {B}ose-{E}instein condensate},\ }\href@noop {} {\bibfield  {journal}
		{\bibinfo  {journal} {\Jprl}\ }\textbf {\bibinfo {volume} {98}},\ \bibinfo
		{pages} {170403} (\bibinfo {year} {2007}{\natexlab{a}})}\BibitemShut
	{NoStop}%
	\bibitem [{\citenamefont {Lugan}\ \emph
		{et~al.}(2007{\natexlab{b}})\citenamefont {Lugan}, \citenamefont {Cl\'ement},
		\citenamefont {Bouyer}, \citenamefont {Aspect},\ and\ \citenamefont
		{Sanchez-Palencia}}]{lugan2007b}%
	\BibitemOpen
	\bibfield  {author} {\bibinfo {author} {\bibfnamefont {P.}~\bibnamefont
			{Lugan}}, \bibinfo {author} {\bibfnamefont {D.}~\bibnamefont {Cl\'ement}},
		\bibinfo {author} {\bibfnamefont {P.}~\bibnamefont {Bouyer}}, \bibinfo
		{author} {\bibfnamefont {A.}~\bibnamefont {Aspect}},\ and\ \bibinfo {author}
		{\bibfnamefont {L.}~\bibnamefont {Sanchez-Palencia}},\ }\bibfield  {title}
	{\bibinfo {title} {{A}nderson localization of {B}ogolyubov quasiparticles in
			interacting {B}ose-{E}instein condensates},\ }\href@noop {} {\bibfield
		{journal} {\bibinfo  {journal} {\Jprl}\ }\textbf {\bibinfo {volume} {99}},\
		\bibinfo {pages} {180402} (\bibinfo {year} {2007}{\natexlab{b}})}\BibitemShut
	{NoStop}%
	\bibitem [{\citenamefont {Gurarie}\ \emph {et~al.}(2009)\citenamefont
		{Gurarie}, \citenamefont {Pollet}, \citenamefont {Prokof'ev}, \citenamefont
		{Svistunov},\ and\ \citenamefont {Troyer}}]{Gurarie2009}%
	\BibitemOpen
	\bibfield  {author} {\bibinfo {author} {\bibfnamefont {V.}~\bibnamefont
			{Gurarie}}, \bibinfo {author} {\bibfnamefont {L.}~\bibnamefont {Pollet}},
		\bibinfo {author} {\bibfnamefont {N.~V.}\ \bibnamefont {Prokof'ev}}, \bibinfo
		{author} {\bibfnamefont {B.~V.}\ \bibnamefont {Svistunov}},\ and\ \bibinfo
		{author} {\bibfnamefont {M.}~\bibnamefont {Troyer}},\ }\bibfield  {title}
	{\bibinfo {title} {Phase diagram of the disordered bose-hubbard model},\
	}\href {https://doi.org/10.1103/PhysRevB.80.214519} {\bibfield  {journal}
		{\bibinfo  {journal} {Phys. Rev. B}\ }\textbf {\bibinfo {volume} {80}},\
		\bibinfo {pages} {214519} (\bibinfo {year} {2009})}\BibitemShut {NoStop}%
	\bibitem [{\citenamefont {S\"oyler}\ \emph {et~al.}(2011)\citenamefont
		{S\"oyler}, \citenamefont {Kiselev}, \citenamefont {Prokof'ev},\ and\
		\citenamefont {Svistunov}}]{Soyler2011}%
	\BibitemOpen
	\bibfield  {author} {\bibinfo {author} {\bibfnamefont {S.~G.}\ \bibnamefont
			{S\"oyler}}, \bibinfo {author} {\bibfnamefont {M.}~\bibnamefont {Kiselev}},
		\bibinfo {author} {\bibfnamefont {N.~V.}\ \bibnamefont {Prokof'ev}},\ and\
		\bibinfo {author} {\bibfnamefont {B.~V.}\ \bibnamefont {Svistunov}},\
	}\bibfield  {title} {\bibinfo {title} {Phase diagram of the commensurate
			two-dimensional disordered {B}ose-{H}ubbard model},\ }\href
	{https://doi.org/10.1103/PhysRevLett.107.185301} {\bibfield  {journal}
		{\bibinfo  {journal} {Phys. Rev. Lett.}\ }\textbf {\bibinfo {volume} {107}},\
		\bibinfo {pages} {185301} (\bibinfo {year} {2011})}\BibitemShut {NoStop}%
	\bibitem [{\citenamefont {Carleo}\ \emph {et~al.}(2013)\citenamefont {Carleo},
		\citenamefont {Bo\'eris}, \citenamefont {Holzmann},\ and\ \citenamefont
		{Sanchez-Palencia}}]{Carleo2013}%
	\BibitemOpen
	\bibfield  {author} {\bibinfo {author} {\bibfnamefont {G.}~\bibnamefont
			{Carleo}}, \bibinfo {author} {\bibfnamefont {G.}~\bibnamefont {Bo\'eris}},
		\bibinfo {author} {\bibfnamefont {M.}~\bibnamefont {Holzmann}},\ and\
		\bibinfo {author} {\bibfnamefont {L.}~\bibnamefont {Sanchez-Palencia}},\
	}\bibfield  {title} {\bibinfo {title} {Universal superfluid transition and
			transport properties of two-dimensional dirty bosons},\ }\href@noop {}
	{\bibfield  {journal} {\bibinfo  {journal} {\Jprl}\ }\textbf {\bibinfo
			{volume} {111}},\ \bibinfo {pages} {050406} (\bibinfo {year}
		{2013})}\BibitemShut {NoStop}%
	\bibitem [{\citenamefont {Sanchez-Palencia}\ and\ \citenamefont
		{Lewenstein}(2010)}]{lsp2010}%
	\BibitemOpen
	\bibfield  {author} {\bibinfo {author} {\bibfnamefont {L.}~\bibnamefont
			{Sanchez-Palencia}}\ and\ \bibinfo {author} {\bibfnamefont {M.}~\bibnamefont
			{Lewenstein}},\ }\bibfield  {title} {\bibinfo {title} {Disordered quantum
			gases under control},\ }\href {https://doi.org/10.1038/nphys1507} {\bibfield
		{journal} {\bibinfo  {journal} {\Jnatphys}\ }\textbf {\bibinfo {volume}
			{6}},\ \bibinfo {pages} {87} (\bibinfo {year} {2010})}\BibitemShut {NoStop}%
	\bibitem [{\citenamefont {Fallani}\ \emph {et~al.}(2007)\citenamefont
		{Fallani}, \citenamefont {Lye}, \citenamefont {Guarrera}, \citenamefont
		{Fort},\ and\ \citenamefont {Inguscio}}]{Fallani2007}%
	\BibitemOpen
	\bibfield  {author} {\bibinfo {author} {\bibfnamefont {L.}~\bibnamefont
			{Fallani}}, \bibinfo {author} {\bibfnamefont {J.~E.}\ \bibnamefont {Lye}},
		\bibinfo {author} {\bibfnamefont {V.}~\bibnamefont {Guarrera}}, \bibinfo
		{author} {\bibfnamefont {C.}~\bibnamefont {Fort}},\ and\ \bibinfo {author}
		{\bibfnamefont {M.}~\bibnamefont {Inguscio}},\ }\bibfield  {title} {\bibinfo
		{title} {Ultracold atoms in a disordered crystal of light: Towards a bose
			glass},\ }\href {https://doi.org/10.1103/PhysRevLett.98.130404} {\bibfield
		{journal} {\bibinfo  {journal} {Phys. Rev. Lett.}\ }\textbf {\bibinfo
			{volume} {98}},\ \bibinfo {pages} {130404} (\bibinfo {year}
		{2007})}\BibitemShut {NoStop}%
	\bibitem [{\citenamefont {Viebahn}\ \emph {et~al.}(2019)\citenamefont
		{Viebahn}, \citenamefont {Sbroscia}, \citenamefont {Carter}, \citenamefont
		{Yu},\ and\ \citenamefont {Schneider}}]{viebahn2019}%
	\BibitemOpen
	\bibfield  {author} {\bibinfo {author} {\bibfnamefont {K.}~\bibnamefont
			{Viebahn}}, \bibinfo {author} {\bibfnamefont {M.}~\bibnamefont {Sbroscia}},
		\bibinfo {author} {\bibfnamefont {E.}~\bibnamefont {Carter}}, \bibinfo
		{author} {\bibfnamefont {J.-C.}\ \bibnamefont {Yu}},\ and\ \bibinfo {author}
		{\bibfnamefont {U.}~\bibnamefont {Schneider}},\ }\bibfield  {title} {\bibinfo
		{title} {Matter-wave diffraction from a quasicrystalline optical lattice},\
	}\href {https://doi.org/10.1103/PhysRevLett.122.110404} {\bibfield  {journal}
		{\bibinfo  {journal} {Phys. Rev. Lett.}\ }\textbf {\bibinfo {volume} {122}},\
		\bibinfo {pages} {110404} (\bibinfo {year} {2019})}\BibitemShut {NoStop}%
	\bibitem [{\citenamefont {Sbroscia}\ \emph {et~al.}(2020)\citenamefont
		{Sbroscia}, \citenamefont {Viebahn}, \citenamefont {Carter}, \citenamefont
		{Yu}, \citenamefont {Gaunt},\ and\ \citenamefont {Schneider}}]{sbroscia2020}%
	\BibitemOpen
	\bibfield  {author} {\bibinfo {author} {\bibfnamefont {M.}~\bibnamefont
			{Sbroscia}}, \bibinfo {author} {\bibfnamefont {K.}~\bibnamefont {Viebahn}},
		\bibinfo {author} {\bibfnamefont {E.}~\bibnamefont {Carter}}, \bibinfo
		{author} {\bibfnamefont {J.-C.}\ \bibnamefont {Yu}}, \bibinfo {author}
		{\bibfnamefont {A.}~\bibnamefont {Gaunt}},\ and\ \bibinfo {author}
		{\bibfnamefont {U.}~\bibnamefont {Schneider}},\ }\bibfield  {title} {\bibinfo
		{title} {Observing localization in a 2{D} quasicrystalline optical lattice},\
	}\href@noop {} {\bibfield  {journal} {\bibinfo  {journal} {Phys. Rev. Lett.}\
		}\textbf {\bibinfo {volume} {125}},\ \bibinfo {pages} {200604} (\bibinfo
		{year} {2020})}\BibitemShut {NoStop}%
	\bibitem [{\citenamefont {Yao}\ \emph {et~al.}(2020)\citenamefont {Yao},
		\citenamefont {Giamarchi},\ and\ \citenamefont {Sanchez-Palencia}}]{Yao2020}%
	\BibitemOpen
	\bibfield  {author} {\bibinfo {author} {\bibfnamefont {H.}~\bibnamefont
			{Yao}}, \bibinfo {author} {\bibfnamefont {T.}~\bibnamefont {Giamarchi}},\
		and\ \bibinfo {author} {\bibfnamefont {L.}~\bibnamefont {Sanchez-Palencia}},\
	}\bibfield  {title} {\bibinfo {title} {Lieb-liniger bosons in a shallow
			quasiperiodic potential: Bose glass phase and fractal mott lobes},\ }\href
	{https://doi.org/10.1103/PhysRevLett.125.060401} {\bibfield  {journal}
		{\bibinfo  {journal} {Phys. Rev. Lett.}\ }\textbf {\bibinfo {volume} {125}},\
		\bibinfo {pages} {060401} (\bibinfo {year} {2020})}\BibitemShut {NoStop}%
	\bibitem [{\citenamefont {Gautier}\ \emph {et~al.}(2021)\citenamefont
		{Gautier}, \citenamefont {Yao},\ and\ \citenamefont
		{Sanchez-Palencia}}]{Gautier2021}%
	\BibitemOpen
	\bibfield  {author} {\bibinfo {author} {\bibfnamefont {R.}~\bibnamefont
			{Gautier}}, \bibinfo {author} {\bibfnamefont {H.}~\bibnamefont {Yao}},\ and\
		\bibinfo {author} {\bibfnamefont {L.}~\bibnamefont {Sanchez-Palencia}},\
	}\bibfield  {title} {\bibinfo {title} {Strongly interacting bosons in a
			two-dimensional quasicrystal lattice},\ }\href
	{https://doi.org/10.1103/PhysRevLett.126.110401} {\bibfield  {journal}
		{\bibinfo  {journal} {Phys. Rev. Lett.}\ }\textbf {\bibinfo {volume} {126}},\
		\bibinfo {pages} {110401} (\bibinfo {year} {2021})}\BibitemShut {NoStop}%
	\bibitem [{\citenamefont {Johnstone}\ \emph {et~al.}(2022)\citenamefont
		{Johnstone}, \citenamefont {Öhberg},\ and\ \citenamefont
		{Duncan}}]{Johnstone_2022}%
	\BibitemOpen
	\bibfield  {author} {\bibinfo {author} {\bibfnamefont {D.}~\bibnamefont
			{Johnstone}}, \bibinfo {author} {\bibfnamefont {P.}~\bibnamefont {Öhberg}},\
		and\ \bibinfo {author} {\bibfnamefont {C.~W.}\ \bibnamefont {Duncan}},\
	}\bibfield  {title} {\bibinfo {title} {Barriers to macroscopic superfluidity
			and insulation in a 2d aubry–andré model},\ }\href
	{https://doi.org/10.1088/1361-6455/ac6d34} {\bibfield  {journal} {\bibinfo
			{journal} {J. Phys. B: At. Mol. Opt. Phys.}\ }\textbf {\bibinfo {volume}
			{55}},\ \bibinfo {pages} {125302} (\bibinfo {year} {2022})}\BibitemShut
	{NoStop}%
	\bibitem [{\citenamefont {Johnstone}\ \emph {et~al.}(2021)\citenamefont
		{Johnstone}, \citenamefont {Öhberg},\ and\ \citenamefont
		{Duncan}}]{Johnstone_2021}%
	\BibitemOpen
	\bibfield  {author} {\bibinfo {author} {\bibfnamefont {D.}~\bibnamefont
			{Johnstone}}, \bibinfo {author} {\bibfnamefont {P.}~\bibnamefont {Öhberg}},\
		and\ \bibinfo {author} {\bibfnamefont {C.~W.}\ \bibnamefont {Duncan}},\
	}\bibfield  {title} {\bibinfo {title} {The mean-field bose glass in
			quasicrystalline systems},\ }\href {https://doi.org/10.1088/1751-8121/ac1dc0}
	{\bibfield  {journal} {\bibinfo  {journal} {J. Phys. A: Math. Theor.}\
		}\textbf {\bibinfo {volume} {54}},\ \bibinfo {pages} {395001} (\bibinfo
		{year} {2021})}\BibitemShut {NoStop}%
	\bibitem [{\citenamefont {Zhu}\ \emph {et~al.}(2023)\citenamefont {Zhu},
		\citenamefont {Yao},\ and\ \citenamefont {Sanchez-Palencia}}]{Zhu2023}%
	\BibitemOpen
	\bibfield  {author} {\bibinfo {author} {\bibfnamefont {Z.}~\bibnamefont
			{Zhu}}, \bibinfo {author} {\bibfnamefont {H.}~\bibnamefont {Yao}},\ and\
		\bibinfo {author} {\bibfnamefont {L.}~\bibnamefont {Sanchez-Palencia}},\
	}\bibfield  {title} {\bibinfo {title} {Thermodynamic phase diagram of
			two-dimensional bosons in a quasicrystal potential},\ }\href
	{https://doi.org/10.1103/PhysRevLett.130.220402} {\bibfield  {journal}
		{\bibinfo  {journal} {Phys. Rev. Lett.}\ }\textbf {\bibinfo {volume} {130}},\
		\bibinfo {pages} {220402} (\bibinfo {year} {2023})}\BibitemShut {NoStop}%
	\bibitem [{\citenamefont {Yu}\ \emph {et~al.}()\citenamefont {Yu},
		\citenamefont {Bhave}, \citenamefont {Reeve}, \citenamefont {Song},\ and\
		\citenamefont {Schneider}}]{JrChiunYu2023}%
	\BibitemOpen
	\bibfield  {author} {\bibinfo {author} {\bibfnamefont {J.-C.}\ \bibnamefont
			{Yu}}, \bibinfo {author} {\bibfnamefont {S.}~\bibnamefont {Bhave}}, \bibinfo
		{author} {\bibfnamefont {L.}~\bibnamefont {Reeve}}, \bibinfo {author}
		{\bibfnamefont {B.}~\bibnamefont {Song}},\ and\ \bibinfo {author}
		{\bibfnamefont {U.}~\bibnamefont {Schneider}},\ }\href@noop {} {\bibinfo
		{title} {Observing the two-dimensional {B}ose glass in an optical
			quasicrystal}},\ \Eprint {https://arxiv.org/abs/ar{X}iv:2303.00737}
	{arXiv:ar{X}iv:2303.00737 [cond-mat.quant-gas]} \BibitemShut {NoStop}%
	\bibitem [{\citenamefont {Gonz\'alez-Tudela}\ and\ \citenamefont
		{Cirac}(2019)}]{cirac2019}%
	\BibitemOpen
	\bibfield  {author} {\bibinfo {author} {\bibfnamefont {A.}~\bibnamefont
			{Gonz\'alez-Tudela}}\ and\ \bibinfo {author} {\bibfnamefont {J.~I.}\
			\bibnamefont {Cirac}},\ }\bibfield  {title} {\bibinfo {title} {Cold atoms in
			twisted-bilayer optical potentials},\ }\href
	{https://doi.org/10.1103/PhysRevA.100.053604} {\bibfield  {journal} {\bibinfo
			{journal} {Phys. Rev. A}\ }\textbf {\bibinfo {volume} {100}},\ \bibinfo
		{pages} {053604} (\bibinfo {year} {2019})}\BibitemShut {NoStop}%
	\bibitem [{\citenamefont {Salamon}\ \emph {et~al.}(2020)\citenamefont
		{Salamon}, \citenamefont {Celi}, \citenamefont {Chhajlany}, \citenamefont
		{Fr\'erot}, \citenamefont {Lewenstein}, \citenamefont {Tarruell},\ and\
		\citenamefont {Rakshit}}]{Salamon2020}%
	\BibitemOpen
	\bibfield  {author} {\bibinfo {author} {\bibfnamefont {T.}~\bibnamefont
			{Salamon}}, \bibinfo {author} {\bibfnamefont {A.}~\bibnamefont {Celi}},
		\bibinfo {author} {\bibfnamefont {R.~W.}\ \bibnamefont {Chhajlany}}, \bibinfo
		{author} {\bibfnamefont {I.}~\bibnamefont {Fr\'erot}}, \bibinfo {author}
		{\bibfnamefont {M.}~\bibnamefont {Lewenstein}}, \bibinfo {author}
		{\bibfnamefont {L.}~\bibnamefont {Tarruell}},\ and\ \bibinfo {author}
		{\bibfnamefont {D.}~\bibnamefont {Rakshit}},\ }\bibfield  {title} {\bibinfo
		{title} {Simulating twistronics without a twist},\ }\href
	{https://doi.org/10.1103/PhysRevLett.125.030504} {\bibfield  {journal}
		{\bibinfo  {journal} {Phys. Rev. Lett.}\ }\textbf {\bibinfo {volume} {125}},\
		\bibinfo {pages} {030504} (\bibinfo {year} {2020})}\BibitemShut {NoStop}%
	\bibitem [{\citenamefont {Luo}\ and\ \citenamefont {Zhang}(2021)}]{Luo2021}%
	\BibitemOpen
	\bibfield  {author} {\bibinfo {author} {\bibfnamefont {X.-W.}\ \bibnamefont
			{Luo}}\ and\ \bibinfo {author} {\bibfnamefont {C.}~\bibnamefont {Zhang}},\
	}\bibfield  {title} {\bibinfo {title} {Spin-twisted optical lattices: Tunable
			flat bands and larkin-ovchinnikov superfluids},\ }\href
	{https://doi.org/10.1103/PhysRevLett.126.103201} {\bibfield  {journal}
		{\bibinfo  {journal} {Phys. Rev. Lett.}\ }\textbf {\bibinfo {volume} {126}},\
		\bibinfo {pages} {103201} (\bibinfo {year} {2021})}\BibitemShut {NoStop}%
	\bibitem [{\citenamefont {Meng}\ \emph {et~al.}(2023)\citenamefont {Meng},
		\citenamefont {Wang}, \citenamefont {Han}, \citenamefont {Liu}, \citenamefont
		{Wen}, \citenamefont {Gao}, \citenamefont {Wang}, \citenamefont {Chin},\ and\
		\citenamefont {Zhang}}]{meng2023atomic}%
	\BibitemOpen
	\bibfield  {author} {\bibinfo {author} {\bibfnamefont {Z.}~\bibnamefont
			{Meng}}, \bibinfo {author} {\bibfnamefont {L.}~\bibnamefont {Wang}}, \bibinfo
		{author} {\bibfnamefont {W.}~\bibnamefont {Han}}, \bibinfo {author}
		{\bibfnamefont {F.}~\bibnamefont {Liu}}, \bibinfo {author} {\bibfnamefont
			{K.}~\bibnamefont {Wen}}, \bibinfo {author} {\bibfnamefont {C.}~\bibnamefont
			{Gao}}, \bibinfo {author} {\bibfnamefont {P.}~\bibnamefont {Wang}}, \bibinfo
		{author} {\bibfnamefont {C.}~\bibnamefont {Chin}},\ and\ \bibinfo {author}
		{\bibfnamefont {J.}~\bibnamefont {Zhang}},\ }\bibfield  {title} {\bibinfo
		{title} {Atomic bose--einstein condensate in twisted-bilayer optical
			lattices},\ }\href {https://doi.org/10.1038/s41586-023-05695-4} {\bibfield
		{journal} {\bibinfo  {journal} {Nature}\ }\textbf {\bibinfo {volume} {615}},\
		\bibinfo {pages} {231} (\bibinfo {year} {2023})}\BibitemShut {NoStop}%
	\bibitem [{\citenamefont {Lee}\ and\ \citenamefont
		{Pixley}(2022)}]{Junhyun2022}%
	\BibitemOpen
	\bibfield  {author} {\bibinfo {author} {\bibfnamefont {J.}~\bibnamefont
			{Lee}}\ and\ \bibinfo {author} {\bibfnamefont {J.~H.}\ \bibnamefont
			{Pixley}},\ }\bibfield  {title} {\bibinfo {title} {Emulating twisted double
			bilayer graphene with a multiorbital optical lattice},\ }\href
	{https://doi.org/10.21468/SciPostPhys.13.2.033} {\bibfield  {journal}
		{\bibinfo  {journal} {SciPost Phys.}\ }\textbf {\bibinfo {volume} {13}},\
		\bibinfo {pages} {033} (\bibinfo {year} {2022})}\BibitemShut {NoStop}%
	\bibitem [{\citenamefont {Paul}\ \emph {et~al.}(2023)\citenamefont {Paul},
		\citenamefont {Recher},\ and\ \citenamefont {Santos}}]{paul2023particle}%
	\BibitemOpen
	\bibfield  {author} {\bibinfo {author} {\bibfnamefont {G.~C.}\ \bibnamefont
			{Paul}}, \bibinfo {author} {\bibfnamefont {P.}~\bibnamefont {Recher}},\ and\
		\bibinfo {author} {\bibfnamefont {L.}~\bibnamefont {Santos}},\ }\bibfield
	{title} {\bibinfo {title} {Particle dynamics and ergodicity-breaking in
			twisted-bilayer optical lattices},\ }\href@noop {} {\bibfield  {journal}
		{\bibinfo  {journal} {arXiv preprint arXiv:2306.01588}\ } (\bibinfo {year}
		{2023})}\BibitemShut {NoStop}%
	\bibitem [{\citenamefont {Yao}\ \emph {et~al.}(2024)\citenamefont {Yao},
		\citenamefont {Tanzi}, \citenamefont {Sanchez-Palencia}, \citenamefont
		{Giamarchi}, \citenamefont {Modugno},\ and\ \citenamefont
		{D'Errico}}]{yao2024}%
	\BibitemOpen
	\bibfield  {author} {\bibinfo {author} {\bibfnamefont {H.}~\bibnamefont
			{Yao}}, \bibinfo {author} {\bibfnamefont {L.}~\bibnamefont {Tanzi}}, \bibinfo
		{author} {\bibfnamefont {L.}~\bibnamefont {Sanchez-Palencia}}, \bibinfo
		{author} {\bibfnamefont {T.}~\bibnamefont {Giamarchi}}, \bibinfo {author}
		{\bibfnamefont {G.}~\bibnamefont {Modugno}},\ and\ \bibinfo {author}
		{\bibfnamefont {C.}~\bibnamefont {D'Errico}},\ }\href@noop {} {\bibinfo
		{title} {Mott transition for a {L}ieb-{L}iniger gas in a shallow
			quasiperiodic potential: {D}elocalization induced by disorder}} (\bibinfo
	{year} {2024}),\ \Eprint {https://arxiv.org/abs/2402.15318} {arXiv:2402.15318
		[cond-mat.quant-gas]} \BibitemShut {NoStop}%
	\bibitem [{\citenamefont {Petrov}\ \emph {et~al.}(2000)\citenamefont {Petrov},
		\citenamefont {Holzmann},\ and\ \citenamefont {Shlyapnikov}}]{Petrov2000}%
	\BibitemOpen
	\bibfield  {author} {\bibinfo {author} {\bibfnamefont {D.~S.}\ \bibnamefont
			{Petrov}}, \bibinfo {author} {\bibfnamefont {M.}~\bibnamefont {Holzmann}},\
		and\ \bibinfo {author} {\bibfnamefont {G.~V.}\ \bibnamefont {Shlyapnikov}},\
	}\bibfield  {title} {\bibinfo {title} {Bose-einstein condensation in quasi-2d
			trapped gases},\ }\href {https://doi.org/10.1103/PhysRevLett.84.2551}
	{\bibfield  {journal} {\bibinfo  {journal} {Phys. Rev. Lett.}\ }\textbf
		{\bibinfo {volume} {84}},\ \bibinfo {pages} {2551} (\bibinfo {year}
		{2000})}\BibitemShut {NoStop}%
	\bibitem [{\citenamefont {Petrov}\ and\ \citenamefont
		{Shlyapnikov}(2001)}]{Petrov2001}%
	\BibitemOpen
	\bibfield  {author} {\bibinfo {author} {\bibfnamefont {D.~S.}\ \bibnamefont
			{Petrov}}\ and\ \bibinfo {author} {\bibfnamefont {G.~V.}\ \bibnamefont
			{Shlyapnikov}},\ }\bibfield  {title} {\bibinfo {title} {Interatomic
			collisions in a tightly confined bose gas},\ }\href
	{https://doi.org/10.1103/PhysRevA.64.012706} {\bibfield  {journal} {\bibinfo
			{journal} {Phys. Rev. A}\ }\textbf {\bibinfo {volume} {64}},\ \bibinfo
		{pages} {012706} (\bibinfo {year} {2001})}\BibitemShut {NoStop}%
	\bibitem [{\citenamefont {Pricoupenko}\ and\ \citenamefont
		{Olshanii}(2007)}]{Pricoupenko_2007}%
	\BibitemOpen
	\bibfield  {author} {\bibinfo {author} {\bibfnamefont {L.}~\bibnamefont
			{Pricoupenko}}\ and\ \bibinfo {author} {\bibfnamefont {M.}~\bibnamefont
			{Olshanii}},\ }\bibfield  {title} {\bibinfo {title} {Stability of
			two-dimensional bose gases in the resonant regime},\ }\href
	{https://doi.org/10.1088/0953-4075/40/11/009} {\bibfield  {journal} {\bibinfo
			{journal} {J. Phys. B: At. Mol. Opt. Phys.}\ }\textbf {\bibinfo {volume}
			{40}},\ \bibinfo {pages} {2065} (\bibinfo {year} {2007})}\BibitemShut
	{NoStop}%
	\bibitem [{\citenamefont {Ha}\ \emph {et~al.}(2013)\citenamefont {Ha},
		\citenamefont {Hung}, \citenamefont {Zhang}, \citenamefont {Eismann},
		\citenamefont {Tung},\ and\ \citenamefont {Chin}}]{Ha2013}%
	\BibitemOpen
	\bibfield  {author} {\bibinfo {author} {\bibfnamefont {L.-C.}\ \bibnamefont
			{Ha}}, \bibinfo {author} {\bibfnamefont {C.-L.}\ \bibnamefont {Hung}},
		\bibinfo {author} {\bibfnamefont {X.}~\bibnamefont {Zhang}}, \bibinfo
		{author} {\bibfnamefont {U.}~\bibnamefont {Eismann}}, \bibinfo {author}
		{\bibfnamefont {S.-K.}\ \bibnamefont {Tung}},\ and\ \bibinfo {author}
		{\bibfnamefont {C.}~\bibnamefont {Chin}},\ }\bibfield  {title} {\bibinfo
		{title} {Strongly interacting two-dimensional bose gases},\ }\href
	{https://doi.org/10.1103/PhysRevLett.110.145302} {\bibfield  {journal}
		{\bibinfo  {journal} {Phys. Rev. Lett.}\ }\textbf {\bibinfo {volume} {110}},\
		\bibinfo {pages} {145302} (\bibinfo {year} {2013})}\BibitemShut {NoStop}%
	\bibitem [{not({\natexlab{a}})}]{note:V1V2}%
	\BibitemOpen
	({\natexlab{a}}),\ \bibinfo {note} {{Further calculations for the unbalanced
			case, $V_1 \neq V_2$, not shown in this work, yield the same phases as
			discussed here and qualitatively similar phase diagrams.}}\BibitemShut
	{Stop}%
	\bibitem [{\citenamefont {Boninsegni}\ \emph
		{et~al.}(2006{\natexlab{a}})\citenamefont {Boninsegni}, \citenamefont
		{Prokof'ev},\ and\ \citenamefont {Svistunov}}]{Boninsegni2006}%
	\BibitemOpen
	\bibfield  {author} {\bibinfo {author} {\bibfnamefont {M.}~\bibnamefont
			{Boninsegni}}, \bibinfo {author} {\bibfnamefont {N.}~\bibnamefont
			{Prokof'ev}},\ and\ \bibinfo {author} {\bibfnamefont {B.}~\bibnamefont
			{Svistunov}},\ }\bibfield  {title} {\bibinfo {title} {Worm algorithm for
			continuous-space path integral monte carlo simulations},\ }\href
	{https://doi.org/10.1103/PhysRevLett.96.070601} {\bibfield  {journal}
		{\bibinfo  {journal} {Phys. Rev. Lett.}\ }\textbf {\bibinfo {volume} {96}},\
		\bibinfo {pages} {070601} (\bibinfo {year} {2006}{\natexlab{a}})}\BibitemShut
	{NoStop}%
	\bibitem [{\citenamefont {Boninsegni}\ \emph
		{et~al.}(2006{\natexlab{b}})\citenamefont {Boninsegni}, \citenamefont
		{Prokof'ev},\ and\ \citenamefont {Svistunov}}]{Boninsegni2006b}%
	\BibitemOpen
	\bibfield  {author} {\bibinfo {author} {\bibfnamefont {M.}~\bibnamefont
			{Boninsegni}}, \bibinfo {author} {\bibfnamefont {N.~V.}\ \bibnamefont
			{Prokof'ev}},\ and\ \bibinfo {author} {\bibfnamefont {B.~V.}\ \bibnamefont
			{Svistunov}},\ }\bibfield  {title} {\bibinfo {title} {Worm algorithm and
			diagrammatic monte carlo: A new approach to continuous-space path integral
			monte carlo simulations},\ }\href
	{https://doi.org/10.1103/PhysRevE.74.036701} {\bibfield  {journal} {\bibinfo
			{journal} {Phys. Rev. E}\ }\textbf {\bibinfo {volume} {74}},\ \bibinfo
		{pages} {036701} (\bibinfo {year} {2006}{\natexlab{b}})}\BibitemShut
	{NoStop}%
	\bibitem [{\citenamefont {Ceperley}(1995)}]{Ceperley1995}%
	\BibitemOpen
	\bibfield  {author} {\bibinfo {author} {\bibfnamefont {D.~M.}\ \bibnamefont
			{Ceperley}},\ }\bibfield  {title} {\bibinfo {title} {Path integrals in the
			theory of condensed helium},\ }\href
	{https://doi.org/10.1103/RevModPhys.67.279} {\bibfield  {journal} {\bibinfo
			{journal} {Rev. Mod. Phys.}\ }\textbf {\bibinfo {volume} {67}},\ \bibinfo
		{pages} {279} (\bibinfo {year} {1995})}\BibitemShut {NoStop}%
	\bibitem [{\citenamefont {Li}\ \emph {et~al.}(2020)\citenamefont {Li},
		\citenamefont {Zhou}, \citenamefont {Mazets}, \citenamefont {Stimming},
		\citenamefont {Møller}, \citenamefont {Zhu}, \citenamefont {Zhai},
		\citenamefont {Xiong}, \citenamefont {Zhou}, \citenamefont {Chen},\ and\
		\citenamefont {Schmiedmayer}}]{Li2020}%
	\BibitemOpen
	\bibfield  {author} {\bibinfo {author} {\bibfnamefont {C.}~\bibnamefont
			{Li}}, \bibinfo {author} {\bibfnamefont {T.}~\bibnamefont {Zhou}}, \bibinfo
		{author} {\bibfnamefont {I.}~\bibnamefont {Mazets}}, \bibinfo {author}
		{\bibfnamefont {H.-P.}\ \bibnamefont {Stimming}}, \bibinfo {author}
		{\bibfnamefont {F.~S.}\ \bibnamefont {Møller}}, \bibinfo {author}
		{\bibfnamefont {Z.}~\bibnamefont {Zhu}}, \bibinfo {author} {\bibfnamefont
			{Y.}~\bibnamefont {Zhai}}, \bibinfo {author} {\bibfnamefont {W.}~\bibnamefont
			{Xiong}}, \bibinfo {author} {\bibfnamefont {X.}~\bibnamefont {Zhou}},
		\bibinfo {author} {\bibfnamefont {X.}~\bibnamefont {Chen}},\ and\ \bibinfo
		{author} {\bibfnamefont {J.}~\bibnamefont {Schmiedmayer}},\ }\bibfield
	{title} {\bibinfo {title} {{Relaxation of bosons in one dimension and the
				onset of dimensional crossover}},\ }\href
	{https://doi.org/10.21468/SciPostPhys.9.4.058} {\bibfield  {journal}
		{\bibinfo  {journal} {SciPost Phys.}\ }\textbf {\bibinfo {volume} {9}},\
		\bibinfo {pages} {058} (\bibinfo {year} {2020})}\BibitemShut {NoStop}%
	\bibitem [{\citenamefont {Fabbri}\ \emph {et~al.}(2011)\citenamefont {Fabbri},
		\citenamefont {Cl\'ement}, \citenamefont {Fallani}, \citenamefont {Fort},\
		and\ \citenamefont {Inguscio}}]{Fabbri2011}%
	\BibitemOpen
	\bibfield  {author} {\bibinfo {author} {\bibfnamefont {N.}~\bibnamefont
			{Fabbri}}, \bibinfo {author} {\bibfnamefont {D.}~\bibnamefont {Cl\'ement}},
		\bibinfo {author} {\bibfnamefont {L.}~\bibnamefont {Fallani}}, \bibinfo
		{author} {\bibfnamefont {C.}~\bibnamefont {Fort}},\ and\ \bibinfo {author}
		{\bibfnamefont {M.}~\bibnamefont {Inguscio}},\ }\bibfield  {title} {\bibinfo
		{title} {Momentum-resolved study of an array of one-dimensional strongly
			phase-fluctuating bose gases},\ }\href
	{https://doi.org/10.1103/PhysRevA.83.031604} {\bibfield  {journal} {\bibinfo
			{journal} {Phys. Rev. A}\ }\textbf {\bibinfo {volume} {83}},\ \bibinfo
		{pages} {031604} (\bibinfo {year} {2011})}\BibitemShut {NoStop}%
	\bibitem [{\citenamefont {Fabbri}\ \emph {et~al.}(2015)\citenamefont {Fabbri},
		\citenamefont {Panfil}, \citenamefont {Cl\'ement}, \citenamefont {Fallani},
		\citenamefont {Inguscio}, \citenamefont {Fort},\ and\ \citenamefont
		{Caux}}]{Fabbri2015}%
	\BibitemOpen
	\bibfield  {author} {\bibinfo {author} {\bibfnamefont {N.}~\bibnamefont
			{Fabbri}}, \bibinfo {author} {\bibfnamefont {M.}~\bibnamefont {Panfil}},
		\bibinfo {author} {\bibfnamefont {D.}~\bibnamefont {Cl\'ement}}, \bibinfo
		{author} {\bibfnamefont {L.}~\bibnamefont {Fallani}}, \bibinfo {author}
		{\bibfnamefont {M.}~\bibnamefont {Inguscio}}, \bibinfo {author}
		{\bibfnamefont {C.}~\bibnamefont {Fort}},\ and\ \bibinfo {author}
		{\bibfnamefont {J.-S.}\ \bibnamefont {Caux}},\ }\bibfield  {title} {\bibinfo
		{title} {Dynamical structure factor of one-dimensional bose gases:
			Experimental signatures of beyond-luttinger-liquid physics},\ }\href
	{https://doi.org/10.1103/PhysRevA.91.043617} {\bibfield  {journal} {\bibinfo
			{journal} {Phys. Rev. A}\ }\textbf {\bibinfo {volume} {91}},\ \bibinfo
		{pages} {043617} (\bibinfo {year} {2015})}\BibitemShut {NoStop}%
	\bibitem [{\citenamefont {Rousseau}\ \emph {et~al.}(2006)\citenamefont
		{Rousseau}, \citenamefont {Arovas}, \citenamefont {Rigol}, \citenamefont
		{H\'ebert}, \citenamefont {Batrouni},\ and\ \citenamefont
		{Scalettar}}]{PhysRevB.73.174516}%
	\BibitemOpen
	\bibfield  {author} {\bibinfo {author} {\bibfnamefont {V.~G.}\ \bibnamefont
			{Rousseau}}, \bibinfo {author} {\bibfnamefont {D.~P.}\ \bibnamefont
			{Arovas}}, \bibinfo {author} {\bibfnamefont {M.}~\bibnamefont {Rigol}},
		\bibinfo {author} {\bibfnamefont {F.}~\bibnamefont {H\'ebert}}, \bibinfo
		{author} {\bibfnamefont {G.~G.}\ \bibnamefont {Batrouni}},\ and\ \bibinfo
		{author} {\bibfnamefont {R.~T.}\ \bibnamefont {Scalettar}},\ }\bibfield
	{title} {\bibinfo {title} {Exact study of the one-dimensional boson hubbard
			model with a superlattice potential},\ }\href
	{https://doi.org/10.1103/PhysRevB.73.174516} {\bibfield  {journal} {\bibinfo
			{journal} {Phys. Rev. B}\ }\textbf {\bibinfo {volume} {73}},\ \bibinfo
		{pages} {174516} (\bibinfo {year} {2006})}\BibitemShut {NoStop}%
	\bibitem [{\citenamefont {Chen}\ \emph {et~al.}(2010)\citenamefont {Chen},
		\citenamefont {Kou}, \citenamefont {Zhang},\ and\ \citenamefont
		{Chen}}]{PhysRevA.81.053608}%
	\BibitemOpen
	\bibfield  {author} {\bibinfo {author} {\bibfnamefont {B.-L.}\ \bibnamefont
			{Chen}}, \bibinfo {author} {\bibfnamefont {S.-P.}\ \bibnamefont {Kou}},
		\bibinfo {author} {\bibfnamefont {Y.}~\bibnamefont {Zhang}},\ and\ \bibinfo
		{author} {\bibfnamefont {S.}~\bibnamefont {Chen}},\ }\bibfield  {title}
	{\bibinfo {title} {Quantum phases of the bose-hubbard model in optical
			superlattices},\ }\href {https://doi.org/10.1103/PhysRevA.81.053608}
	{\bibfield  {journal} {\bibinfo  {journal} {Phys. Rev. A}\ }\textbf {\bibinfo
			{volume} {81}},\ \bibinfo {pages} {053608} (\bibinfo {year}
		{2010})}\BibitemShut {NoStop}%
	\bibitem [{\citenamefont {Roux}\ \emph {et~al.}(2008)\citenamefont {Roux},
		\citenamefont {Barthel}, \citenamefont {McCulloch}, \citenamefont {Kollath},
		\citenamefont {Schollw\"ock},\ and\ \citenamefont
		{Giamarchi}}]{PhysRevA.78.023628}%
	\BibitemOpen
	\bibfield  {author} {\bibinfo {author} {\bibfnamefont {G.}~\bibnamefont
			{Roux}}, \bibinfo {author} {\bibfnamefont {T.}~\bibnamefont {Barthel}},
		\bibinfo {author} {\bibfnamefont {I.~P.}\ \bibnamefont {McCulloch}}, \bibinfo
		{author} {\bibfnamefont {C.}~\bibnamefont {Kollath}}, \bibinfo {author}
		{\bibfnamefont {U.}~\bibnamefont {Schollw\"ock}},\ and\ \bibinfo {author}
		{\bibfnamefont {T.}~\bibnamefont {Giamarchi}},\ }\bibfield  {title} {\bibinfo
		{title} {Quasiperiodic bose-hubbard model and localization in one-dimensional
			cold atomic gases},\ }\href {https://doi.org/10.1103/PhysRevA.78.023628}
	{\bibfield  {journal} {\bibinfo  {journal} {Phys. Rev. A}\ }\textbf {\bibinfo
			{volume} {78}},\ \bibinfo {pages} {023628} (\bibinfo {year}
		{2008})}\BibitemShut {NoStop}%
	\bibitem [{\citenamefont {Zhu}\ \emph {et~al.}(2024)\citenamefont {Zhu},
		\citenamefont {Yu}, \citenamefont {Johnstone},\ and\ \citenamefont
		{Sanchez-Palencia}}]{Zhu2024}%
	\BibitemOpen
	\bibfield  {author} {\bibinfo {author} {\bibfnamefont {Z.}~\bibnamefont
			{Zhu}}, \bibinfo {author} {\bibfnamefont {S.}~\bibnamefont {Yu}}, \bibinfo
		{author} {\bibfnamefont {D.}~\bibnamefont {Johnstone}},\ and\ \bibinfo
		{author} {\bibfnamefont {L.}~\bibnamefont {Sanchez-Palencia}},\ }\bibfield
	{title} {\bibinfo {title} {Localization and spectral structure in
			two-dimensional quasicrystal potentials},\ }\href
	{https://doi.org/10.1103/PhysRevA.109.013314} {\bibfield  {journal} {\bibinfo
			{journal} {Phys. Rev. A}\ }\textbf {\bibinfo {volume} {109}},\ \bibinfo
		{pages} {013314} (\bibinfo {year} {2024})}\BibitemShut {NoStop}%
	\bibitem [{not({\natexlab{b}})}]{note:SupplMat}%
	\BibitemOpen
	({\natexlab{b}}),\ \bibinfo {note} {{See details in the Supplemental
			Material. It discusses the extraction of tunnelling rates from effective
			tight-binding models and the finite temperature behaviour of the SF
			domains.}}\BibitemShut {Stop}%
	\bibitem [{not({\natexlab{c}})}]{note:rt}%
	\BibitemOpen
	({\natexlab{c}}),\ \bibinfo {note} {{Note that the global four-fold rotation
			symmetry is, however, preserved since $J^\beta_{b;x} =J^\beta_{c;y}$ and
			$J^\beta_{b;y} =J^\beta_{c;x}$.}}\BibitemShut {Stop}%
	\bibitem [{\citenamefont {Gerbier}(2007)}]{Gerbier2007}%
	\BibitemOpen
	\bibfield  {author} {\bibinfo {author} {\bibfnamefont {F.}~\bibnamefont
			{Gerbier}},\ }\bibfield  {title} {\bibinfo {title} {Boson mott insulators at
			finite temperatures},\ }\href {https://doi.org/10.1103/PhysRevLett.99.120405}
	{\bibfield  {journal} {\bibinfo  {journal} {Phys. Rev. Lett.}\ }\textbf
		{\bibinfo {volume} {99}},\ \bibinfo {pages} {120405} (\bibinfo {year}
		{2007})}\BibitemShut {NoStop}%
	\bibitem [{\citenamefont {Troyer}\ \emph {et~al.}(1998)\citenamefont {Troyer},
		\citenamefont {Ammon},\ and\ \citenamefont {Heeb}}]{troyer1998}%
	\BibitemOpen
	\bibfield  {author} {\bibinfo {author} {\bibfnamefont {M.}~\bibnamefont
			{Troyer}}, \bibinfo {author} {\bibfnamefont {B.}~\bibnamefont {Ammon}},\ and\
		\bibinfo {author} {\bibfnamefont {E.}~\bibnamefont {Heeb}},\ }\bibfield
	{title} {\bibinfo {title} {Parallel object oriented {M}onte {C}arlo
			simulations},\ }\href@noop {} {\bibfield  {journal} {\bibinfo  {journal}
			{Lect. Notes Comput. Sci.}\ }\textbf {\bibinfo {volume} {1505}},\ \bibinfo
		{pages} {191} (\bibinfo {year} {1998})}\BibitemShut {NoStop}%
	\bibitem [{\citenamefont {Bauer}\ \emph {et~al.}(2011)\citenamefont {Bauer},
		\citenamefont {Carr}, \citenamefont {Evertz}, \citenamefont {Feiguin},
		\citenamefont {Freire}, \citenamefont {Fuchs}, \citenamefont {Gamper},
		\citenamefont {Gukelberger}, \citenamefont {Gull}, \citenamefont {Guertler},
		\citenamefont {Hehn}, \citenamefont {Igarashi}, \citenamefont {Isakov},
		\citenamefont {Koop}, \citenamefont {Ma}, \citenamefont {Mates},
		\citenamefont {Matsuo}, \citenamefont {Parcollet}, \citenamefont {Pawlowski},
		\citenamefont {Picon}, \citenamefont {Pollet}, \citenamefont {Santos},
		\citenamefont {Scarola}, \citenamefont {Schollwöck}, \citenamefont {Silva},
		\citenamefont {Surer}, \citenamefont {Todo}, \citenamefont {Trebst},
		\citenamefont {Troyer}, \citenamefont {Wall}, \citenamefont {Werner},\ and\
		\citenamefont {Wessel}}]{ALPS2011}%
	\BibitemOpen
	\bibfield  {author} {\bibinfo {author} {\bibfnamefont {B.}~\bibnamefont
			{Bauer}}, \bibinfo {author} {\bibfnamefont {L.~D.}\ \bibnamefont {Carr}},
		\bibinfo {author} {\bibfnamefont {H.}~\bibnamefont {Evertz}}, \bibinfo
		{author} {\bibfnamefont {A.}~\bibnamefont {Feiguin}}, \bibinfo {author}
		{\bibfnamefont {J.}~\bibnamefont {Freire}}, \bibinfo {author} {\bibfnamefont
			{S.}~\bibnamefont {Fuchs}}, \bibinfo {author} {\bibfnamefont
			{L.}~\bibnamefont {Gamper}}, \bibinfo {author} {\bibfnamefont
			{J.}~\bibnamefont {Gukelberger}}, \bibinfo {author} {\bibfnamefont
			{E.}~\bibnamefont {Gull}}, \bibinfo {author} {\bibfnamefont {S.}~\bibnamefont
			{Guertler}}, \bibinfo {author} {\bibfnamefont {A.}~\bibnamefont {Hehn}},
		\bibinfo {author} {\bibfnamefont {R.}~\bibnamefont {Igarashi}}, \bibinfo
		{author} {\bibfnamefont {S.}~\bibnamefont {Isakov}}, \bibinfo {author}
		{\bibfnamefont {D.}~\bibnamefont {Koop}}, \bibinfo {author} {\bibfnamefont
			{P.}~\bibnamefont {Ma}}, \bibinfo {author} {\bibfnamefont {P.}~\bibnamefont
			{Mates}}, \bibinfo {author} {\bibfnamefont {H.}~\bibnamefont {Matsuo}},
		\bibinfo {author} {\bibfnamefont {O.}~\bibnamefont {Parcollet}}, \bibinfo
		{author} {\bibfnamefont {G.}~\bibnamefont {Pawlowski}}, \bibinfo {author}
		{\bibfnamefont {J.}~\bibnamefont {Picon}}, \bibinfo {author} {\bibfnamefont
			{L.}~\bibnamefont {Pollet}}, \bibinfo {author} {\bibfnamefont
			{E.}~\bibnamefont {Santos}}, \bibinfo {author} {\bibfnamefont
			{V.}~\bibnamefont {Scarola}}, \bibinfo {author} {\bibfnamefont
			{U.}~\bibnamefont {Schollwöck}}, \bibinfo {author} {\bibfnamefont
			{C.}~\bibnamefont {Silva}}, \bibinfo {author} {\bibfnamefont
			{B.}~\bibnamefont {Surer}}, \bibinfo {author} {\bibfnamefont
			{S.}~\bibnamefont {Todo}}, \bibinfo {author} {\bibfnamefont {S.}~\bibnamefont
			{Trebst}}, \bibinfo {author} {\bibfnamefont {M.}~\bibnamefont {Troyer}},
		\bibinfo {author} {\bibfnamefont {M.}~\bibnamefont {Wall}}, \bibinfo {author}
		{\bibfnamefont {P.}~\bibnamefont {Werner}},\ and\ \bibinfo {author}
		{\bibfnamefont {S.}~\bibnamefont {Wessel}},\ }\bibfield  {title} {\bibinfo
		{title} {The {ALPS} project release 2.0: Open source software for strongly
			correlated systems},\ }\href@noop {} {\bibfield  {journal} {\bibinfo
			{journal} {J. Stat. Mech.: Th. Exp.}\ }\textbf {\bibinfo {volume} {05}},\
		\bibinfo {pages} {P05001} (\bibinfo {year} {2011})}\BibitemShut {NoStop}%
\end{thebibliography}

%

\renewcommand{\theequation}{S\arabic{equation}}
\setcounter{equation}{0}
\renewcommand{\thefigure}{S\arabic{figure}}
\setcounter{figure}{0}
\renewcommand{\thesection}{S\arabic{section}}
\setcounter{section}{0}
\onecolumngrid  


\newpage

{\center \bf \large Supplemental Material for \\}
{\center \bf \large Weak Superfluidity in Twisted Optical Potentials \\ \vspace*{1.cm}
}

This supplemental material gives details about
the construction of band-dependent tight-binding models and the extraction of effective tunnelling energies (Sec. I)
and the finite temperature behaviour of the SF domains (Sec. II).
\appendix

\section{I. Band-dependent effective tight-binding models} \label{sc_app2}

\begin{figure*}[b!]
	\centering
	\makebox[0pt]{\includegraphics[width=0.9\linewidth]{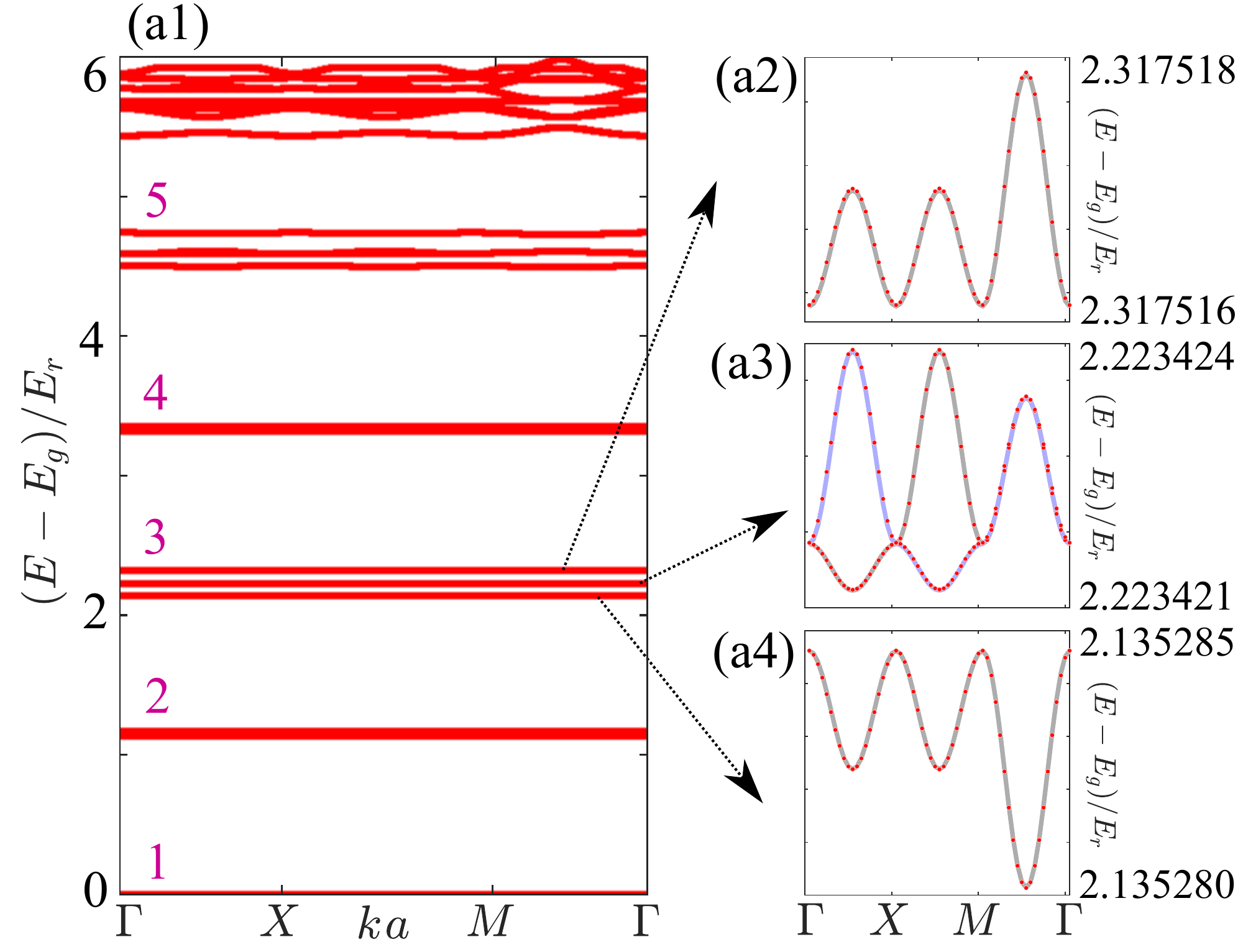}}
	\caption{
		(a1)~Low-energy single-particle spectrum for the commensurate potential with $\theta=\theta_{3,5} \approx 28.07^{\circ}$ and $V=6\Er$, along high symmetry points of the first Brillouin zone. 
		(a2)-(a4)~Zooms in to the subbands of band $\beta=3$.
	}
	\label{fig_bands1}
\end{figure*}

To determine the tunnelling energies in periodic moir\'e potentials, we compare the energy spectra as found using continuous-space exact diagonalization of the single-particle Hamiltonian and the prediction of effective tight-binding models.
On the one hand, we calculate the exact dispersion relations $\varepsilon(\mathbf{k})$, with $\mathbf{k}$ the quasi-momentum, by applying the Bloch transform $\psi (\mathbf{r})=e^{i\mathbf{k} \cdot \mathbf{r}} u(\mathbf{r})$ and solving the Schrödinger-like equation for $u(\mathbf{r})$ within a single unit cell, with length $L=\ell_{3,5}$. Figure~\ref{fig_bands1}(a1) shows the single-particle spectrum $\varepsilon(\mathbf{k})$ hence obtained for the commensurate potential with $V=6\Er$, across the high symmetry points of the first Brillouin zone, i.e.~$\Gamma \rightarrow  \mathbf{k} = (-\pi/\ell_{3,5},-\pi/\ell_{3,5})$, $X \rightarrow  \mathbf{k} = (+\pi/\ell_{3,5},-\pi/\ell_{3,5})$ and $M \rightarrow  \mathbf{k} = (+\pi/\ell_{3,5},+\pi/\ell_{3,5})$. For sufficiently large potential depths $V$, the single-particle spectrum splits into distinct energy bands, where eigenstates fill degenerate local minima of the potential consecutively, see for instance Figs. 2(a3) and (a4) of the main paper.
The ground-state band corresponds to the filling of the degenerate global minima arranged in a simple square lattice and displays the standard cosine-like dispersion of Eq. (6) in the main paper, the width of which yields the tunnelling energy, as discussed in the main paper.

The other bands split into distinct subbands.  Each one displays a cosine-like dispersion relation, although some of them are distorted, see zooms given for band~3 in Figs.~\ref{fig_bands1}(a2)-(a4).
To understand this, we note that the eigenstates now correspond to the filling of 4 local minima within each unit cell.
To model this structure in band $\beta$, we first separate tunnelling energies between sites/spots into two distinct energy scales: tunnellings within a unit cell, noted with an $I$, and tunnellings between adjacent unit cells, noted as $J_\alpha$, where
$\alpha$ labels the subband index (the band index $\beta$ is omitted to simplify notations).
In general, intra-cell couplings are much larger than inter-cell couplings, which allows us to treat the latter in perturbation of the former.
Within a unique cell, we simply have a 4 sites, with an example in Fig. 2(a4) of the main paper for band 3.
Taking $I$ as the nearest-neighbour tunnelling and $I'$ as the next-nearest-neighbour tunnelling, the Hamiltonian can be written as
\begin{equation} \label{eq_h_lt}
\hat{H} = 
\begin{pmatrix}
\epsilon & -I & -I' & -I \\
-I & \epsilon & -I & -I' \\
-I'  & -I  & \epsilon & -I  \\
-I & -I' & -I & \epsilon
\end{pmatrix},
\end{equation}
with $\epsilon$ the on-site energy.
This matrix has eigenvalues
\begin{equation}
\begin{aligned}
E_a & = \epsilon + 2I -I', \\
E_{B,C} & = \epsilon + I', \\
E_d & = \epsilon -2I -I',
\end{aligned}
\end{equation}
with degeneracy $E_B=E_C$.
To capture tunnelling between sites in different unit cells, we then treat the problem as a lattice of coupled 4-level systems, arranged as a square lattice with period $\ell_{3,5}$. The 4-level system corresponds to the 4 eigenstates of the matrix in Eq.~\eqref{eq_h_lt} and the inter-cell couplings are level-dependent. Energy-separation between the subbands allows us to restrict to couplings between
equal energy eigenstates. It allows us to write an effective Hamiltonian of the form of
\begin{equation}
\begin{aligned}
\hat{H}  =
& \, \, E_a\sum_i \hat{a}_i^\dagger \hat{a}_i - J_a \sum_{\langle i,j\rangle} \hat{a}_i^\dagger \hat{a}_j
+ E_d\sum_i \hat{d}_i^\dagger \hat{d}_i - J_d \sum_{\langle i,j\rangle} \hat{d}_i^\dagger \hat{d}_j
\\ 
& + E_B\sum_i (\hat{B}_i^\dagger \hat{B}_i + \hat{C}_i^\dagger \hat{C}_i)
-J_B \sum_{\langle i,j\rangle} ( \hat{B}_i^\dagger \hat{B}_j + \hat{C}_i^\dagger \hat{C}_j ) 
-\sum_{\langle i,j\rangle} J'_{B;i,j} ( \hat{B}_i^\dagger \hat{C}_j + \hat{C}_i^\dagger \hat{B}_j ),
\end{aligned}
\end{equation}
where $i$ and $j$ span the lattice sites,
$\langle i,j\rangle$ denotes nearest-neighbour summations,
and the operators $\hat{a}_i$, $\hat{B}_i$, $\hat{C}_i$, and $\hat{d}_i$
annihilate a particle in the corresponding eigenstate of Eq.~\eqref{eq_h_lt}.
Note that the coupling $J'_{B;i,j}$ between the two distinct states $\ket{B}$ and $\ket{C}$ depends on the direction, owing to the symmetry breaking induced by the choice of those states
but, due to $4$-fold rotational symmetry, we necessary have $J'_{B;x}$ = -$J'_{B;y} \equiv J'_B$, with $x$ and $y$ denoting the tunnelling across the $x$ and $y$ axis.
By transforming the operators to momentum space, we then have
\begin{equation} \label{eq_h_ft}
\begin{aligned}
\hat{H} = 
\sum_\mathbf{k} \varepsilon_a(\mathbf{k}) \hat{a}_\mathbf{k}^\dagger \hat{a}_\mathbf{k}
+ \sum_\mathbf{k} \varepsilon_d(\mathbf{k}) \hat{d}_\mathbf{k}^\dagger \hat{d}_\mathbf{k}
+
\sum_\mathbf{k} \varepsilon_B(\mathbf{k}) ( \hat{B}_\mathbf{k}^\dagger \hat{B}_\mathbf{k}
+ \hat{C}_\mathbf{k}^\dagger \hat{C}_\mathbf{k} )
+ \sum_\mathbf{k} \varepsilon'_B(\mathbf{k}) (\hat{C}_\mathbf{k}^\dagger \hat{B}_\mathbf{k} + \hat{B}_\mathbf{k}^\dagger \hat{C}_\mathbf{k}),
\end{aligned}
\end{equation}
where
\begin{equation}
\begin{aligned}
\varepsilon_a(\mathbf{k}) & = E_a - 2J_a \left[\cos (k_x \ell_{3,5}) + \cos (k_y \ell_{3,5})\right],
\\ \varepsilon_B(\mathbf{k}) & = E_B - 2J_B \left[\cos (k_x \ell_{3,5}) + \cos(k_y \ell_{3,5})\right],
\\ \varepsilon'_B(\mathbf{k}) & =  - 2J'_B \left[\cos (k_x \ell_{3,5}) - \cos(k_y \ell_{3,5})\right],
\\ \varepsilon_d(\mathbf{k}) & = E_d - 2J_d \left[\cos (k_x \ell_{3,5}) + \cos (k_y \ell_{3,5})\right].
\end{aligned}
\end{equation}
We may then diagonalise the summation involving $\hat{B}_\mathbf{k}$ and $\hat{C}_\mathbf{k}$ operator introducing the new operators
\begin{equation}
\begin{aligned}
\hat{b}_\mathbf{k} = ( \hat{B}_\mathbf{k} + \hat{C}_\mathbf{k} )/\sqrt{2}
\qquad \textrm{and} \qquad
\hat{c}_\mathbf{k} = ( \hat{B}_\mathbf{k} - \hat{C}_\mathbf{k} )/\sqrt{2},
\end{aligned}
\end{equation}
which allows us to rewrite Eq.~\eqref{eq_h_ft} as
\begin{equation}
\begin{aligned}
\hat{H} = 
& \sum_\mathbf{k} \varepsilon_a(\mathbf{k}) \hat{a}_\mathbf{k}^\dagger \hat{a}_\mathbf{k}
+ \sum_\mathbf{k} \varepsilon_d(\mathbf{k}) \hat{d}_\mathbf{k}^\dagger \hat{d}_\mathbf{k}
+
\sum_\mathbf{k} \varepsilon_b(\mathbf{k}) \hat{b}_\mathbf{k}^\dagger \hat{b}_\mathbf{k} + \sum_\mathbf{k} \varepsilon_c(\mathbf{k}) \hat{c}_\mathbf{k}^\dagger \hat{c}_\mathbf{k},
\end{aligned}
\end{equation}
where
\begin{equation}
\begin{aligned}
\varepsilon_b(\mathbf{k}) & = E_B - 2\big[ J_b \cos (k_x \ell_{3,5}) + J_c \cos (k_y \ell_{3,5}) \big],\\
\varepsilon_c(\mathbf{k}) & =  E_B - 2\big[ J_c \cos (k_x \ell_{3,5}) + J_b \cos (k_y \ell_{3,5}) \big].
\end{aligned}
\end{equation}
and
\begin{equation}
J_b = J_B + J'_B
\qquad \textrm{and} \qquad J_c=J_B - J'_B.
\end{equation}
Note that, while each band $b$ and $c$ breaks the 4-fold rotation symmetry, the set of both restores it.

Finally, by fitting $\varepsilon_a(\mathbf{k})$, $\varepsilon_b(\mathbf{k})$, $\varepsilon_c(\mathbf{k})$, and $\varepsilon_d(\mathbf{k})$ to the continuous dispersion relations of each subband found from diagonalization of the continuous-space Schrödinger equation, we are then able to determine all of the inter-cell couplings $J$ and $J'$ of the effective lattice model for different bands $\beta$.
In Table~\ref{table_tunnellings}, we give the largest ($J^{\beta}_{\textrm{max}}$) and smallest ($J^{\beta}_{\textrm{min}}$) inter-cell couplings hence found for each band.

\begin{table}[h!]
	\centering
	\begin{tabular}{||c c c||} 
		\hline
		\textbf{Band} & $J^{\beta}_{\textrm{min}}/\Er$ & $J^{\beta}_{\textrm{max}}/\Er$ \\ [0.5ex] 
		\hline\hline
		\textbf{1} & $4.40 \times 10^{-11}$  & $4.40 \times 10^{-11}$ \\ 
		\textbf{2} & $8.34 \times 10^{-10}$ & $1.24 \times 10^{-6}$ \\
		\textbf{3} & $7.89 \times 10^{-7}$ & $1.12 \times 10^{-6}$ \\		
		\textbf{4} & $2.84 \times 10^{-4}$ & $2.91 \times 10^{-4}$ \\
		\textbf{5} & $2.70 \times 10^{-3}$ & $4.09 \times 10^{-3}$ \\ [1ex] 
		\hline
	\end{tabular}
	\caption{Tunnelling coefficients of the effective lattice models for the five lowest-energy bands of the commensurate potential with twist angle $\theta_{3,5}$ and amplitude $V=6\Er$.}
	\label{table_tunnellings}
\end{table}

\section{II. Finite temperature behaviour of weak superfluid domains} \label{sc_app1}
\begin{figure}[t!]
	\centering
	\makebox[0pt]{\includegraphics[width=0.9\linewidth]{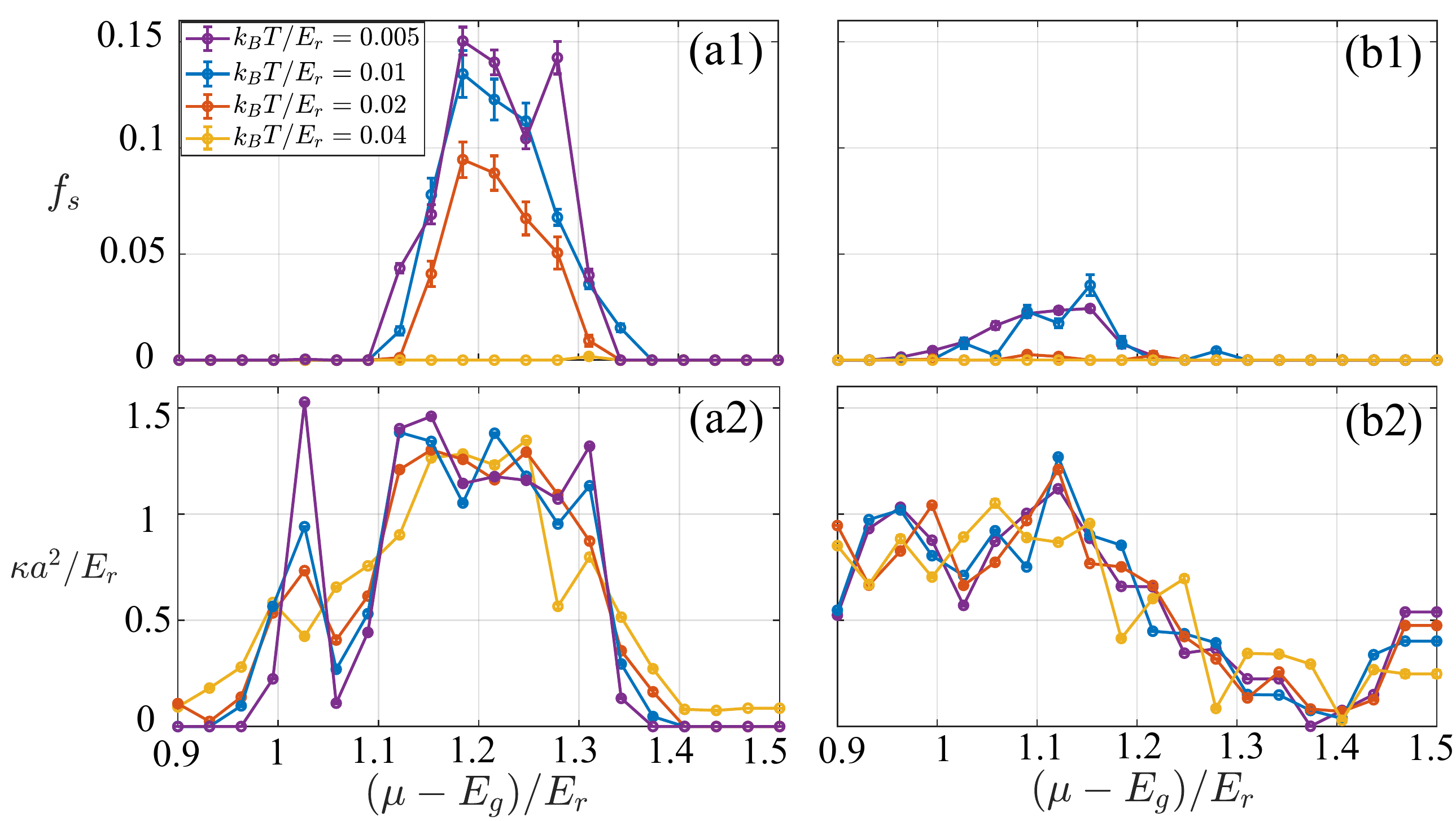}}
	\caption{(a1)-(b1) Superfluid fraction $\fs$ and (a2)-(b2) compressibility $\kappa$ across a small range of $\mu$ for the same parameters as Figs.~2(a2) and (b2) of the main paper when $V=3\Er$, for the commensurate (a1)-(a2) and incommensurate (b1)-(b2) angle respectively.
	}
	\label{fig_tmpC}
\end{figure}

Here, we show results for the behaviour of weak SF domains as a function of temperature.
{In Fig.~\ref{fig_tmpC}, we plot the (a1)~superfluid fraction $\fs$ and (a2)~compressibility $\kappa$ across the phase diagram for the commensurate
	otential (Fig. 2(a2) from the main paper), for the fixed potential amplitude $V=3\Er$.}
As expected, we find that a sizeable $\fs$ appears roughly when $\kB  T$ is of the order of $J^{2}_{\textrm{\tiny max}}\approx 0.0134\Er$ (i.e.~$\kB T/\Er \approx 0.03 \pm 0.01$).
Starting with the larger temperature of $T=0.04\Er/\kB$ (yellow curve) in Figs.~\ref{fig_tmpC}(a1)-(a2), we find a broad region with $\fs=0$ and $\kappa > 0$, clearly indicating the presence of a NF.
By decreasing the temperature, we find
that a SF domain appears with increasing $\fs$, slowly growing in width across $\mu$.
Furthermore, towards the left and right hand sides of Fig.~\ref{fig_tmpC}(a2), insulating regions, in which $\kappa=0$, are also stabilised
when the temperature decreases,
corresponding to the onset of DW plateaus.
The remaining NF domains in which $\kappa>0$ and $\fs=0$ therefore become smaller for decreasing $\kB T/\Er$, as expected.
Near by $\mu-E_g \approx 1$, we find a peak in compressibility. We expect that it becomes a SF but only at even lower temperatures.

The counterparts for the incommensurate potential (corresponding to Fig. 2(b2) from the main paper) is shown in Figs.~\ref{fig_tmpC}(b1)-(b2). Here we observe different behaviour: The SF fraction $\fs$ remains vanishingly small, except in a small domain where it is finite for the smaller temperatures $T\le0.01\Er/\kB$, with values an order of magnitude smaller than that for the commensurate case. The compressibility is non-zero at each point, and the regions with $\fs=0$ and $\kappa>0$ correspond to a BG.

\end{document}